\documentclass[10pt,journal,compsoc]{IEEEtran}
\IEEEoverridecommandlockouts

\usepackage{amsmath,amssymb,amsfonts}
\usepackage{algorithmic}
\usepackage{graphicx}
\usepackage{textcomp}
\usepackage{xcolor}
\usepackage{hyperref}
\usepackage{cleveref}
\usepackage{xspace}
\usepackage{paralist}
\usepackage{multirow}
\usepackage{subcaption}
\usepackage{tikz}
\usepackage{courier}

\definecolor{Gray}{gray}{0.3}
\tikzstyle{mybox} = [draw=black, very thick, rectangle, rounded corners, inner ysep=5pt, inner xsep=5pt, fill=gray!20]

\newcommand{\takeaway}[2]{
    \smallskip
    \noindent
    \begin{tikzpicture}
        \node [mybox] (box){%
        \centering
        \begin{minipage}{.465\textwidth}
        \fontsize{8.8}{10}\selectfont
        \textbf{Finding #1}. #2
        \end{minipage}
        };
    \end{tikzpicture}%
}

\newcommand{\approach}{{\sc Funcre}\xspace}
\newcommand{\etal}{\emph{et al.}\xspace}
\newcommand{\ie}{\emph{i.e.},\xspace}
\newcommand{\eg}{\emph{e.g.},\xspace}

\newcommand{\bgs}{{\sc BugSwarm}\xspace}
\newcommand{\bds}{{\sc BuildSwarm}\xspace}
\newcommand{\unpara}[1]{\smallskip \noindent \underline {\bf #1:}}
\usepackage{array}
\usepackage{graphics}
\usepackage{lipsum}

\newcolumntype{L}[1]{>{\raggedright\let\newline\\\arraybackslash\hspace{0pt}}m{#1}}
\newcolumntype{C}[1]{>{\centering\let\newline\\\arraybackslash\hspace{0pt}}m{#1}}
\newcolumntype{R}[1]{>{\raggedleft\let\newline\\\arraybackslash\hspace{0pt}}m{#1}}

\ifCLASSOPTIONcompsoc
  \usepackage[nocompress]{cite}
\else
  \usepackage{cite}
\fi
\ifCLASSINFOpdf
\else
\fi
\begin{document}
%
\title{Learning to Find Usages of Library Functions in Optimized Binaries}
%
%
%
%

\author{Toufique~Ahmed,
        Premkumar~Devanbu,
        and~Anand~Ashok~Sawant
\IEEEcompsocitemizethanks{\IEEEcompsocthanksitem All the authors are with the Department
of Computer Science, University of California, Davis,
CA, 95616.\protect\\
E-mail: \{tfahmed, ptdevanbu, asawant\}@ucdavis.edu }
\thanks{Manuscript in submission}}

%
%

\markboth{IEEE Transactions on Software Engineering}%
{Shell \MakeLowercase{\textit{et al.}}: Bare Demo of IEEEtran.cls for Computer Society Journals}
%



\IEEEtitleabstractindextext{%
\begin{abstract}
Much software, whether beneficent or malevolent, is distributed only as binaries, sans source code. 
Absent source code, understanding binaries' behavior can be quite challenging,  especially when compiled under higher levels of compiler optimization. 
These optimizations can transform comprehensible, ``natural" source constructions into something entirely unrecognizable. 
Reverse engineering binaries, especially those suspected of being malevolent or guilty of intellectual property theft, are important and time-consuming tasks. There is a great deal of interest in tools to ``decompile" binaries back into more natural source code to
aid reverse engineering. 
Decompilation involves several desirable steps, including
recreating source-language constructions, variable names, and perhaps even comments. 
One central step in creating binaries is  optimizing function calls, using steps such as inlining. 
Recovering these (possibly inlined) function calls from optimized binaries is an essential task that most state-of-the-art decompiler tools try to do but do not perform very well. 
In this paper, we evaluate a supervised learning approach to the problem of recovering  optimized function calls. We leverage open-source software and develop an automated labeling scheme to generate a reasonably large
dataset of binaries labeled with actual function usages. 
We augment this large but limited labeled dataset with a pre-training step, which learns the decompiled code statistics from a much larger unlabeled dataset.
Thus augmented, our learned labeling model can be combined with an existing decompilation tool, Ghidra, to achieve substantially improved performance in function call recovery, especially at higher levels
of optimization.

\end{abstract}

}

\maketitle

\IEEEdisplaynontitleabstractindextext

%
\IEEEpeerreviewmaketitle


\IEEEraisesectionheading{\section{Introduction}\label{sec:introduction}}
\IEEEPARstart{I}{n} their seminal work, Chikofsky and Cross~\cite{chikofsky1990reverse}, 
define \emph{Software Reverse Engineering} as ``the process of analyzing a subject system to (1) identify the system's components and their interrelationships and (2) create representations of the system in another form or at a higher level of abstraction''. Understanding the behavior of software binaries generated by potentially
untrusted parties have many motivations such binaries may incorporate
 \eg stolen intellectual property (such as patented or protected algorithms or data), unauthorized access to 
system resources, or malicious behavior of various sorts. 
The ability to reverse engineer binaries would make it more difficult to conceal
potential bad behavior, and thus act as a deterrent, and this would
enhance public confidence in, and overall free exchange of, software technology. 

Reverse engineering of an \emph{optimized} binary is quite challenging, 
since compilers substantially transform source code structures to create the binary. This is done primarily to improve run-time performance;  however, some compilers support deliberate obfuscation of source code details, to protect IP, or for security reasons. The resulting binary could be stripped of all information such as variable names, code comments, and user-defined types. Binaries compiled using \verb+gcc+ can be optimized in the interest of run-time performance benefits, (even if compilation \emph{per se} takes longer).  
Optimizations include function inlining, array vectorization, and loop unrolling. This dramatically alters the code at the assembly level, making it substantially more challenging to decompile the binary successfully.

Two tools are industry standard for reverse engineering: Hexrays IDA Pro~\cite{hexrays} and Ghidra~\cite{ghidra}. These tools incorporate two distinct functionalities:  a \emph{disassembler} (convert binary to assembly code), and a \emph{decompiler} (convert assembly code to a higher-order representation, similar to C code), and a variety of data flow analysis tools. Both tools can handle binaries that have been compiled on a variety of architectures (such as x86/64 or ARM64). 

\begin{figure*}[h!]
    \centering
    \includegraphics[width=\textwidth]{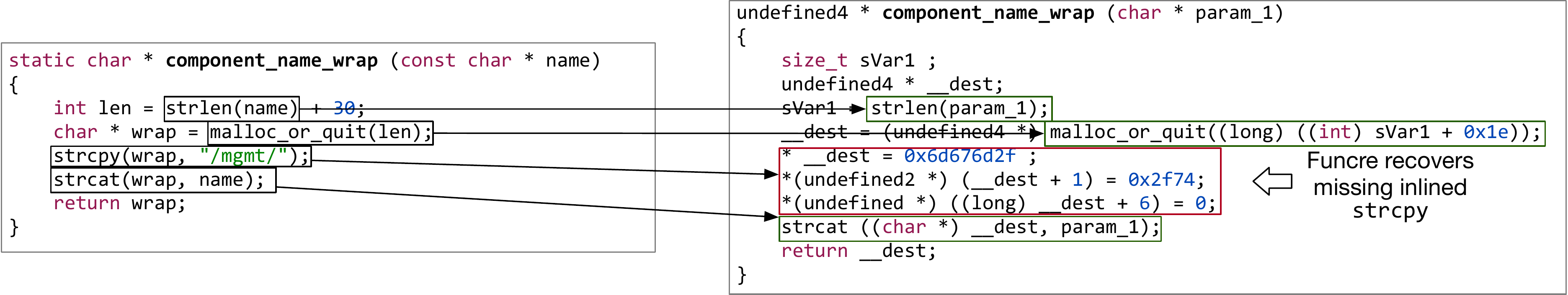}
    \caption[Caption]{Finding inline functions in real-world\protect\footnotemark decompiled version of original C source code}
    
    \label{fig:find}
\end{figure*}

In the field of security, quite a bit of work has focused on understanding the behavior of malicious applications by examining their library API calls~\cite{schultz2000data,wang2006surveillance,ye2007imds,ye2010cimds,ye2010hierarchical,ye2009intelligent,masud2008mining,tian2010differentiating,islam2013classification}. The intuition behind this is that calls made to library APIs (such as Windows DLLs) can capture the important underlying semantics of the malware's attacking behaviour~\cite{ye2017survey}. However, uncovering API calls is particularly hard as the compiler might have mangled the inlined body of the \emph{called} function together with the code at the \emph{calling} site in complex ways. Both Ghidra and Hexrays have specially
engineered functionality for recovering calls to library functions . 
These functions are considered very important by the developers of these tools and are explicitly documented
and advertised.

Both solutions use a database of assembly-level signatures for each potentially inlined library call. Assembler code recovered from the binary is matched against the database to identify library function bodies within the binary. The approaches, however, do not work as well with higher optimization levels due to extensive code restructuring~\cite{flirt,qiu2015library} and context-dependent inlining of the called method's body. Despite both tools making an attempt to find library function invocations in a static way (see, \emph{e.g.,}~\cite{hexrays-inline}, item 7), the problem is non-trivial and more often than not these functions are not recovered. This problem is complicated by varying compilers, compiler versions, optimization levels, and library versions.

This signature or pattern-oriented approach adopted by current tools  relies on a pattern database; 
this database must be \emph{manually maintained}  to include a pattern (or patterns) for each possible inlined function. Therefore, we adopt a more data-driven approach and propose to \emph{learn} to identify those functions that are most often used in the data. To that end, we develop our tool, \approach, which finds inlined library function invocations even in binaries compiled with higher optimization levels. With sufficient data, a 
suitably designed training approach, and a powerful enough model, this approach offers enhanced ability to recover inlined functions. 
As an illustration, in~\Cref{fig:find}, we present a sample of original source code (from GitHub) 
and its Ghidra output for compilation with Os. From the decompiled version,  it's evident that recovering {\small\tt strlen}, {\small\tt malloc\_or\_quit} and, {\small\tt strcat} is possible for Ghidra, 
but the {\small\tt strcpy} gets inlined with copying of pointers to stack addresses  after optimization. The pattern-database approach used by Ghidra works well for the three of the four functions; 
but Ghidra fails to recover the latter, (\emph{strcpy}), because the more complex pattern required is not available in its database. 
The precision of Ghidra is 1.0 for this example, but the recall is lower (0.75). Our tool
\approach builds on Ghidra, and will only try to recover additional inlined functions beyond what Ghidra does, using
\emph{automatically learned patterns} latent within its layered deep-learning architecture. In this specific case, \approach is able to recover the {\small\tt strcpy}, thereby improving Ghidra's recall to 1.0.
\approach, can do this because it can \emph{learn} to find the
patterns (within decompiled code) that reflect the occurrence of the various possible inlining patterns of functions
within the decompiled source of optimized binaries.

We note that our tool, \approach, \emph{starts with}, and  \emph{builds upon} the output of Ghidra decompiler. It works on top of Ghidra-decompiled source code 
output; it tries to recover \emph{only those}  functions missed by Ghidra.   
Our training dataset consists of \emph{decompiled}  binaries (built at scale by compiling open-source code) with labels indicating
the position of inlined function calls contained within, specifically those inlined functions which the decompiler \emph{failed}
to recover.  
We created a build-pipeline that leverages Docker and Travis CI; we successfully built binaries for almost 1,200 projects. We then decompile these binaries using Ghidra to obtain a learned vector representation wherefrom we recover inlined functions. To obtain labeled data (\ie knowing which function has been inlined where), we develop a custom labeling approach that allows us to places labels (with the name of the function) in the binary, at locations where a function has been inlined. Our labels \emph{resist} being removed by optimizing compilers; this also minimizes interference with the optimizing compilers code generation. This labeling approach allows us to successfully annotate 9,643 files and 459 open-source projects, creating a large training dataset.

Our approach uses a pre-training plus fine-tuning scheme~\cite{liu2019roberta} currently popular in natural language processing. 
We pre-train using a masked language model (MLM) on decompiled binaries. This pre-training step helps our model
learn the statistics of a large corpus of decompiled source code. These statistics are represented in a low-dimensional
positional embedding. 
We then fine-tune on the labeled dataset of inlined functions. \approach achieves a precision of 0.64 and a recall of 0.46. When combined with the results of Ghidra (since it uses the output of Ghidra, which might contain some functions recovered), our approach provides the highest f-score across all optimization levels while also recovering a larger number of unique functions. Recovering inlined functions (where assembly-level function prologues \& epilogues are absent, the inlined body is changed and shuffled by optimization) is hard even for humans and existing tools; however, a high-capacity neural model (RoBERTa + Transformer) that uses pre-training and fine-tuning can still learn statistical features and associations between inlined functions and the mangled residual clues that signal their presence. This enables the recovery of 19\% more unique functions over Ghidra and 10\% over Hexrays.
We make the following contributions: 

\begin{enumerate}
\item 
We have generated a training set of 670K fragments for fine-tuning for inline function recovery from 8.6K binaries, each having at least one inline function. We also have a validation set of 25K fragments (from 270 binaries) and a separate test set for each optimization level. 

\item 
We pre-train a RoBERTa~\cite{liu2019roberta}architecture with an enormous amount of unlabeled decompiled binary data. We fine-tune it for the downstream task inline function recovery and achieve state-of-the-art performance. We also show that Transformers with pre-trained embedding works better than standard Transformers proposed by Vaswani \emph{et al.}~\cite{vaswani2017attention}.  

\item
We improve the F-score of library function recovery by  3\%-12\% at different optimization levels without introducing a significant amount of false positives. 
\end{enumerate}

\footnotetext{Original source from GitHub: \url{https://github.com/rlite/rlite/blob/aaec531682756f17c36249e71013a5d3b4f374f9/user/tools/rlite-ctl.c\#L1064}}

\section{Background}
\subsection{Importance of function calls in binary comprehension}
Eisenbarth~\etal~\cite{eisenbarth2001aiding} and Schultz~\etal~\cite{schultz2000data} argue that identifying library function calls made within a binary are key to comprehending its behavior. Substantial prior research in the field of binary analysis has focused on this problem. 

Much of the effort to understand
binaries is to identify malware. Schultz~\etal\cite{schultz2000data} use the calls made to Windows system APIs to understand if a binary has malicious intentions. Ye~\etal\cite{ye2007imds} found that reverse engineers can identify malware in a binary based on the co-occurrence of six calls made to a specific kernel API. 

A barrier to static analysis techniques is that sometimes binaries can be optimized and/or obfuscated. To overcome this, researchers have used dynamic analysis to understand the APIs being accessed by a binary. Hunt and Brubacher~\cite{hunt1999detours} and Willems~\etal~\cite{willems2007toward} attempt to detect calls made to Windows system APIs by instrumenting the system libraries. Bayer~\etal~\cite{bayer2006ttanalyze} and Song~\etal~\cite{song2008bitblaze} emulate the Windows runtime and recover the Windows system API calls. In comparison to static analysis, dynamic analysis is limited by test set coverage, 
as well as by dynamic cloaking (malware could disguise its behavior when it knows it being surveilled, \emph{e.g.,} in a VM).

Given how important library/API calls are to reverse engineers' understanding of the semantics of a binary, it is pivotal that these calls are recovered by disassembler and decompiler. However, optimizing compilers can inline many library calls, thereby making them hard to recover, even by state-of-the-art tooling; improving library function recovery is an important problem.

\subsection{Disassembler vs decompiler}
\label{sec:dis_v_dec}
Binaries are created in two stages: \begin{inparaenum}[(1)]\item source code is pre-processed and compiled into machine code and \item the machine code is linked with all supporting code such as libraries and system calls to create the executable binary.\end{inparaenum} 
\xspace Similarly, the process of reverse engineering of a binary comprises of two stages: \begin{inparaenum}[(1)]
    \item the binary is ``disassembled'' into assembler code and
    \item the assembler code is converted into a higher-order representation which is close to the original source code.
\end{inparaenum}

\emph{Disassemblers} such as Ghidra, Binary Ninja, IDA Pro, and gdb perform the first stage of reverse engineering. Since the machine instructions in a binary generally have a one-to-one mapping with the assembly instructions for most platforms, disassembly \emph{per se} is relatively straightforward.

Next, \emph{decompilers} such as Ghidra, Hex-rays, and Snowman transform the machine code produced by the disassembler into a more legible representation referred to as pseudo-C code. Decompilers produce code that is ostensibly more compact and readable than that produced by a disassembler. They also often recover the original function 
boundaries (\emph{i.e.}, where function \emph{bodies} start and end), control flow structure, and primitive type information. This makes decompilers a better tool for a reverse engineer as the effort and knowledge required to understand pseudo-C code is less than that of reading assembler code.

\subsection{Library function recovery}
\label{sec:inline}
Given its importance, we focus on the recovery of function calls from binaries. The two main reverse engineering tools - Hex-rays IDA Pro (commercially available with a license cost of approximately \$10,000\footnote{Estimate based on the cost of base IDA Pro disassembler license and the cost adding three platform-specific Hexrays decompilers}) and Ghidra (open source and maintained by the National Security Agency) - have dedicated solutions targeted just for recovering function calls. The developers of Hexrays acknowledge the importance of function recovery by stating that: 
``Sometimes, the knowledge of the class of a library function can considerably ease the analysis of a program. This knowledge might be extremely helpful in discarding useless information.''~\cite{flirt}

Both tools can identify function calls. 
They maintain a database of function signatures at the byte level (assembler code). They recover the function by checking each sequence or block of operations in the disassembled code against the signatures in the database.

We observe that majority of the previous research work in this field is based on call graph matching which has been designed to be robust to statement interleaving due to compiled optimization. These approaches are static in nature and try to go beyond the offerings of Ghidra and Hex-rays.

Qiu \etal~\cite{qiu2015library,qiu2015using} implement a static approach to recover inlined library functions by extracting execution dependence graphs for a library function and then matching this in decompiled code to recover. This work reports deal with inlined functions in optimized binaries, however, the evaluation lacks a performance breakdown by optimization level. Furthermore, only precision numbers are reported on a small subset of inlined string library functions, and the overall performance is not compared to Ghidra or Hex-Rays.

BinShape by Shirani~\etal~\cite{shirani2017binshape} also uses graph features for function identification. However, they do not assess the efficacy of their approach against inlined functions. ``impact of inlined functions" were not scrutinized.  We are the first to attempt this task with a neural model and compare this to the SOTA tools such as Ghidra and HexRays.

There have been a couple of neural approaches to recovering function names but not inlined functions. He \etal~\cite{he2018debin} present a tool called Debin that is based on a combination of a decision-tree-based classification algorithm and a probabilistic graph model that attempts to recover function invocations and other symbols in obfuscated code. David \etal~\cite{david2020neural} encode the control flow graphs of invocation sites and try to recover the correct invocation name using LSTM's and Transformers. Neither approach explicitly deals with inlined library functions nor present any results broken down by optimization level.

\begin{figure*}
    \centering
    \includegraphics[width=\textwidth]{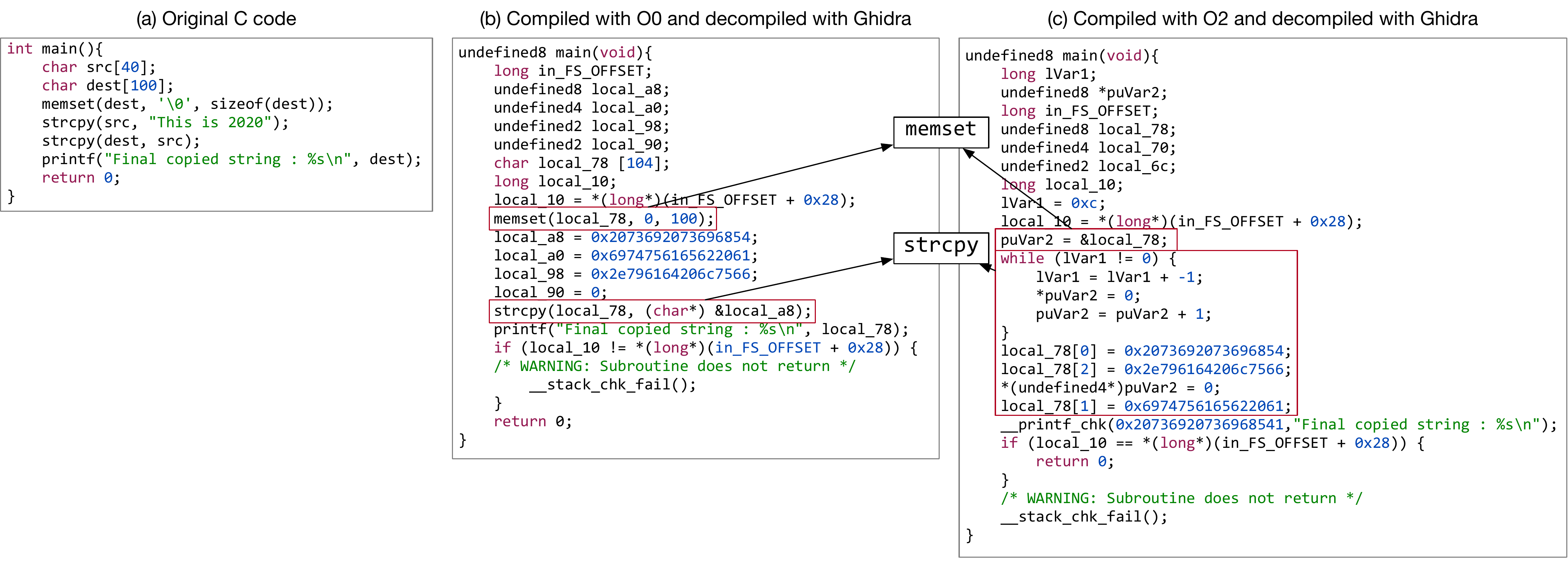}
    \caption{Comparison between original source code (a), Ghidra output for compilation with O0 (b) and Ghidra output for compilation with O2 (c)\label{fig:compare}}
\end{figure*}

The function recovery in these tools has a major flaw: they aren't good at recovering functions from binaries that have been compiled with optimizations. In C and C++, there are six optimization levels (in increasing order of complexity: O0, O1, Os, O2, O3, and, Of). Code compiled with O0 is the most basic: compilation is fast, and no optimization is done. Starting from O1, more and more optimizations are introduced, and the code structure changes substantially.  
At Of, the most optimized level, the compiler does not even guarantee correctness. Hex-rays IDA Pro and Ghidra work better
 with code that has been compiled using the O0 or O1 optimization levels, since the code structure is largely preserved. 

In the toy example seen in \Cref{fig:compare}(a), we see that the source code of a file is written in C. The function depicted invokes \verb+memset+ and then \verb+strcpy+ (twice). When we compile this file with no optimizations (the O0 flag) and then decompile it using Ghidra (output seen in \Cref{fig:compare}(b)), we see that the decompiler can recover called functions and can create a generally good representation of the original C code. Note, however, that it ``fuses'' the chained string copy invocations. When we compile with a higher optimization level such as O2 and then decompile the file, we see the result in \Cref{fig:compare}(c). 
The performance of the decompiler degrades, as the binary gets more optimized: some library function uses are no longer recovered. In the figure, we highlight the parts of the function relating to the library function calls,  whose implementation has been inlined. In this example, we do see that three other function calls are recovered, however, this could be due to the fact that they were never inlined or that Ghidra does a good job of recovering them even if they were inlined.

We want to clarify that like~\cite{bao2014byteweight, wang2017semantics, shin2015recognizing, pei2020xda} we do not target the function boundary identification task. Ghidra \& Hexrays already do this at 90\%+ accuracy. They do much worse at the recovery of inlined library functions; This task is a challenge for the heuristic method used by Ghidra \& Hexrays, especially at higher optimization levels, as acknowledged by the developers~\cite{hexrays}; by leveraging powerful neural models (explained in Section~\ref{sec:model}), \approach can improve these tools.


A decompiler can recover most of the semantics of the original code, however, it has a hard time recovering variable names, struct types, exact data flow, code comments, and inlined library functions. Most of this information is lost in the compilation - decompilation loop, \eg in \Cref{fig:compare} the decompiler adds a lot of new local variables each with an ambiguous name.

State-of-the-art approaches to improving decompilation employ machine learning (ML) techniques. Katz~\etal~\cite{katz2019towards} propose to decompile disassembled code by using a Neural Machine Translation (NMT) model. Their approach currently works at the statement level and seeks to recover natural C code for each block of disassembled code. Lacomis~\etal~\cite{lacomis2019dire} use an encoder-decoder model to recover variable names. Their approach only targets code compiled with O0 and not on higher optimizations.

We see that ML approaches to improving decompilation are limited. We hypothesize that an ML-based approach will work well for the task of library function recovery because ML can detect patterns in highly unstructured data.

\section{Approach/Methodology}

\subsection{Research questions}

Our approach to improving library function recovery builds on top of the pseudo-C code produced by the Ghidra decompiler. 
Using large volumes of source-available projects and some careful, automated instrumentation, we develop a \emph{supervised learning} approach
to find library functions that other tools are unable to find. Our first RQ considers the effectiveness of our approach \ie how effective is \approach at recovering library function invocations not recovered by Ghidra.

\noindent\textbf{RQ1a: }\emph{How effective is our supervised learning-based approach in recovering library function usage?}

In Natural Language Processing (NLP), it is now well-established that pre-training a high-capacity model
using self-supervision for an artificial task (\eg predicting a deleted token, or the following
sentence) improves performance on practical tasks like question-answering.  
Pre-training forces the  layers of the model to learn position-dependent
embeddings of tokens that efficiently capture the statistics of token co-occurrence in very large corpora. 
These embeddings are a very useful, transferable representation of a token and its context~\cite{devlin2018bert,liu2019roberta} which
substantially improves performance on other tasks. By using them as an initial embedding within
a (possibly different) task-specific network and ``fine-tuning" using data labeled specifically for that task, 
much higher performance can be achieved, even if the task-specific labeled data is limited. The benefits
of this approach have also been reported for code~\cite{feng2020codebert,kanade2019pre}. We examine
whether pre-training over large corpora of decompiled pseudo-C can be helpful for our task of recovering library function invocations not recovered by Ghidra. 

\noindent\textbf{RQ1b: }\emph{How much does pre-training with ROBERTa help with library function usage recovery?}

C and C++ binaries can be compiled with a variety of optimizations (see \Cref{sec:inline}). Most disassemblers and decompilers can handle code with no optimizations. In line with that, past research that uses a deep learning (DL) model also targets code compiled with no optimizations. However, in our work, we target higher optimization levels as well. We assess the performance of our model on five optimization levels:

\noindent\textbf{RQ2: }\emph{How does optimization level affect the performance of \approach?}

With machine-learning approaches, the training data can strongly influence the test results. The model might perform better on library functions more prevalent in training data. 

\noindent\textbf{RQ3: }\emph{How does the popularity of library methods influence test results?}

Finally, we assess whether our model outperforms current tools when it comes to retrieving library functions in decompiled code:

\noindent\textbf{RQ4: }\emph{How does \approach perform in relation to state-of-the-art approaches?}


\subsection{Dataset creation}

\begin{figure*}
    \centering
    \includegraphics[scale=0.42, trim={0 6cm 0 0}]{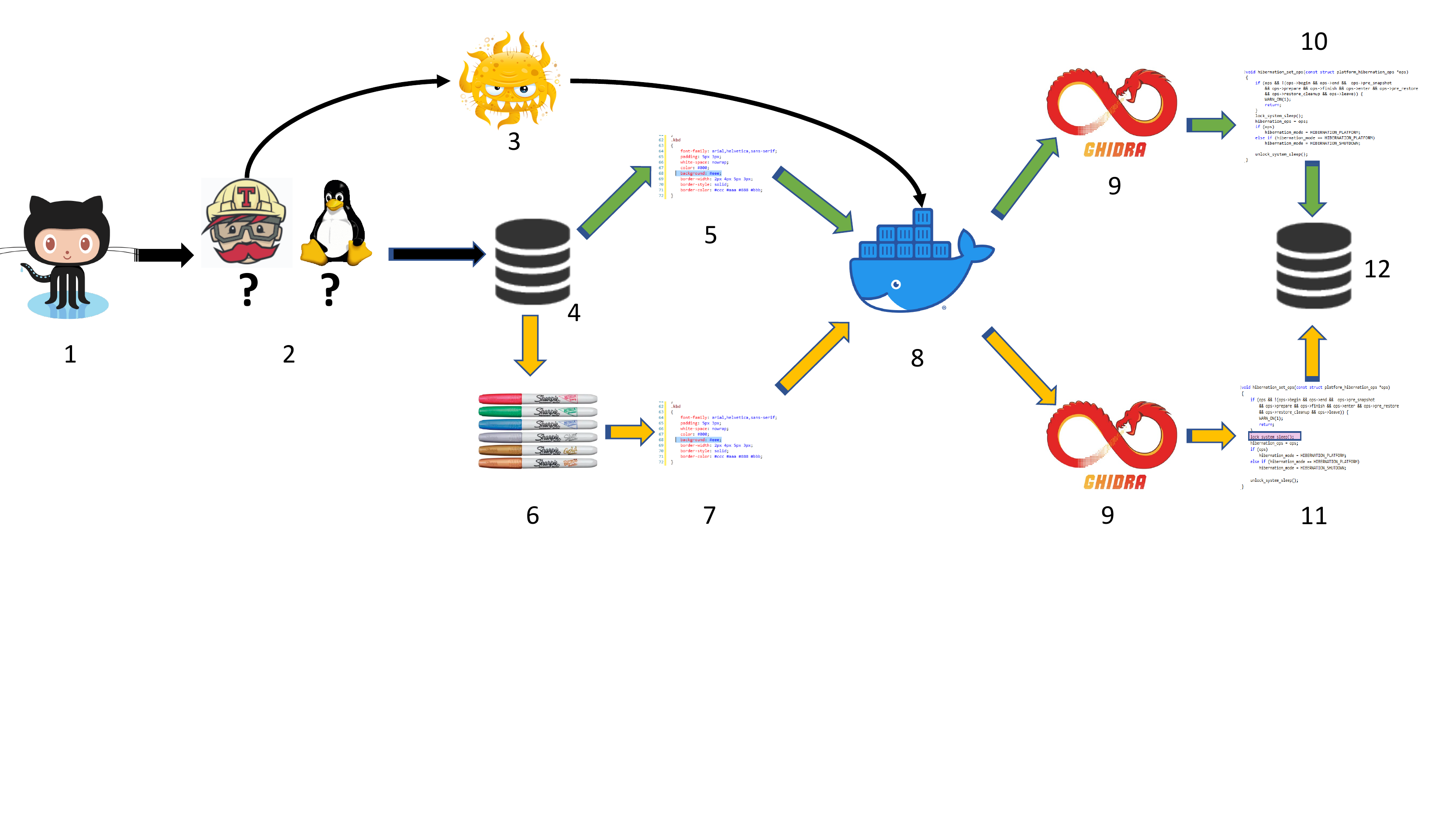}
    \caption{Training data pipeline. We mine projects from GitHub (1); after filtering (2) the ones enabled for Travis, and certain Operating systems, the projects are gathered in a source dataset (4). We then adapt the publicly available BugSwarm toolset (3) to mine Docker containers (8) for building. We indelibly instrument (6)
    the library function invocations in the source to get marked source code (7). The raw (5) and marked (7) sources are built using the Docker containers (8); we then use
    Ghidra (9) to decompile matched pairs (10,11) of marked and unmarked decompiled sources, which are gathered into our labeled dataset (12). \label{fig:data_pipeline}} 
\end{figure*}
A key requirement for using supervised machine learning for library function recovery is the creation of a curated, labeled
dataset where the occurrence of in-lined functions within decompiled binaries is labeled. There is currently no such dataset of labeled decompiled C/C++ binaries, and we have had to create our own. 
This presented several challenges.

\begin{enumerate}
\item
\emph{Large scale, diverse data.} We need a broadly representative, large dataset that captures relevant statistics of current coding styles, library/API usages, compiler settings, and platforms. 
\item
\emph{Reproducible Builds.} To create binaries with \emph{labeled} inlined library functions we need to suitably instrument the source to insert labels, and then
reproduce the build procedures of a large, diverse set of projects. Build procedures are notoriously brittle, with many tricky dependencies, and so challenging to reproduce~\cite{tomassi2019bugswarm}. 
\item
\emph{Indelible labels.} Because optimizing compilers make substantial changes to the logic of the code, our approach to creating binaries where the original inlined library functions could be labeled in a way that endures after optimization \& decompilation is a tricky business. 
\end{enumerate}

We employ a multi-stage project selection and build process to meet these challenges (an overview of which can be seen in \Cref{fig:data_pipeline}) as elucidated below:

\unpara{Large-scale, Diverse data}
The focus of this work is to recover library functions from C-based binaries. Since modern deep-learning models are ``data-hungry'', we need the largest possible corpus of built binaries aligned with its original C source code. We sourced data from GitHub. Our selection criteria for projects is as follows:

\begin{enumerate}
    \item \emph{Projects under active development.} We determine a project's activity by checking for commits in the previous six months (as of April 2020). This helps ensure that selected projects are representative of current coding styles and API usages. 
    \item \emph{Projects using Travis as their CI build platform.} We select those with public build histories and logs (hosted on \url{travis.org}) so that we can replicate the builds.
    \item \emph{Projects with available Build Containers.} We filter out projects that declare in their Travis configuration (.travis.yml file) that their platform requirement is either Mac OS/X or Linux version 12.04. Travis does not provide open-source Docker containers for either build platform, thus making a build irreproducible.
\end{enumerate}

Our initial selection comprised 10,000 actively developed C-based projects. After filtering for Travis-based projects and then on their build distribution, we are left with 2,634 projects.

\unpara{Reproducible Builds\label{sec:building}}\xspace Successfully re-building projects at scale requires the downloading of each project's dependencies and ensuring that the correct build platform is used. This is a frustrating, failure-prone process under the best of circumstances. These problems are exacerbated when building C projects as there is no standard dependency management system comparable to those in languages such as Java (Maven or Gradle), Python (pip) and, C\# (NuGet).

All 2,634 projects in our filtered set of GitHub-based C projects use Travis for continuous integration and require one of three Linux distributions: Trusty (14.04), Xenial (16.04), or Bionic (18.04). Travis CI builds each project in a clean docker container: it first installs all required dependencies and then invokes build and test scripts. We aim to recreate this process.

Fortunately, we were able to leverage the open-source \bgs~\cite{tomassi2019bugswarm} infrastructure. \bgs was originally developed to reproduce buildable pairs of buggy and fixed versions of large, real-world systems. To ensure reliable builds and reproducible failures, the \bgs pipeline builds pairs of commits and tests them five times. For our purposes, we do not need pairs; we just need reproducible, 
test-passing (not failing), singleton builds. \bgs is able to identify the right Docker container that a project uses, download the project into the container, install dependencies and build the project using its scripts and the Travis configuration of the project. We only need this part of the pipeline that can build just the latest commit of the code. 
We downloaded the source-available \bgs~\cite{bugswarm} project and adapted it for our purposes. 
First, \bgs currently does not support Travis builds for languages other than Java and Python. We augment BugSwarm's capability to deal with C-based projects. Second, we refactored the \bgs code to retain only a single, buildable version from the latest version of active projects. This adapted version of \bgs called ``\bds" (which we will make available upon publication of this work~\cite{replication}) works as follows. 

\begin{enumerate}
    \item For each project, we use the Travis API, to download a list of public Travis builds; along with each build, we also download its details, such as build configuration, date of the build, and associated jobs for the build.
    \item From this list of builds, we select the latest passing build. Each build might have more than one job associated with it~\cite{docker-matrix} For this build, we select the first job that suits our criteria \begin{inparaenum}[(1)]
    \item the job fits our OS criteria (see above), \item the job does not require Docker as a service (some projects require child Docker containers for testing, a scenario we cannot reproduce)
    to build the project and \item the job needs either a \verb+gcc+ or \verb+clang+ compiler.
    \end{inparaenum} 
    \item For the selected job, we create the Travis build script that can replicate the entire build procedure, using the Travis build utility~\cite{travis-build}.
    \item From the downloaded log for the job, we parse the Docker image that was used by Travis. Travis releases most of their docker images on docker hub~\cite{travis-docker}. We use the same image as our base image to build the project and add the build script to it by composing a docker image.
    \item Once this docker image is built, we run the docker build script (generated earlier) on the project inside a docker container. This build script downloads the dependencies builds the code to produce binaries.
    \item If the project builds successfully in the container, we release the docker image to Dockerhub~\cite{binswarm-docker}, and retain that image tag so that the image can be reused; we also collect the pair of the C source file and its object file.
\end{enumerate}

\noindent\underline{{\emph {Disassembling and decompiling a binary}}}
For each project that we can re-build, we need to decompile its binary \ie convert the executable into a form of pseudo-C code. \Cref{sec:dis_v_dec} explains the process of disassembling and decompiling the binary to recover the pseudo-C code.

The two main tools for disassembling and decompiling a binary are Ghidra and Hexrays IDA Pro. We select Ghidra as our base tool, as it is open source and is freely available; however, we also baseline an evaluation set against both tools,
 for the specific task of identifying inlined library functions.

Ghidra can disassemble and decompile an executable (.exe file). This entails separating the executable into smaller object files (.o files). Each of these object files is then disassembled by delinking the libraries that they depend on, and then the resulting assembler code is decompiled. In our case, we directly operate on the object files and not on the full executable. This is because we have a one-to-one mapping between the object file and its corresponding C source file. This results in us creating a dataset with source code, binary code, and decompiled code triplets.


\unpara{Indelible Labels}
\label{sec:marking}
Our machine-learner must learn to associate patterns within the decompiled code with specific
inlined library functions. 
To provide a supervisory signal for the learner, we must
\emph{identify} and \emph{locate} a in-lined functions within the Pseudo C code produced by the decompiler; this  is non-trivial.
It's difficult to align the original C source code with the decompiled Pseudo C, since optimization, followed by the decompilation process, can mangle the original code beyond recognition.  
Thus, recovering a one-to-one mapping between the original code and the decompiled code is virtually impossible (especially when compiled using higher optimization levels). 

To create our dataset of decompiled code with labeled locations for inlined library functions, we need to inject some form of \emph{robust label} into the original C source, that would survive despite optimization transformations and be
inserted into the binary; this could then be recovered by a decompiler. We refer to this system of annotating inlined library functions in the binary as \emph{indelible labeling}, or ``marking" for short. 

The process of marking starts by injecting a marker header in each function in a C project and each function from the libraries used by the project. For our purposes, we wish to train a learner to learn to identify \emph{just} those functions \emph{not} identified by
current decompilers. To find these, we first compile and then decompile source files. 
For those inlined library functions which are \emph{not recovered} by the decompiler, 
we must insert a marker that indicates the name of the function and its location within the decompiled Pseudo C  code. 
The marking process must meet several requirements: 
the injected marker must not interfere with the compilation and decompilation process (no errors should be triggered). 
Second, there must be no inadvertent changes to the final compiled binary that would differentiate the marked binary from the unmarked: 
if the resulting decompiled marked Pseudo C differs too much from the original Pseudo C, the training signal for our learner
would be confused, and the marked inline library function would not be reliably recoverable by the learner.  
Third, the injected marker must be resistant to change or removal by the compiler's optimization process. For example,
if we were to insert a marker, as a code fragment, whose results were not used elsewhere in the code, for example, something naive like:

{\footnotesize\tt char *marker1 = "function\_inlined: printf()";}

\noindent then the compiler might, for example, find that {\small\tt marker1} is not used elsewhere, and just simply remove the marking statement, thus
robbing us of a data label. We tried several approaches to this problem: 

\begin{enumerate}
    \item \verb+printf+: Injecting a statement that prints the name of the function being inlined. We found that a \verb+printf+ statement could at times change the structure of the Ghidra output. For lower optimization levels, the code does not change much; however, for higher optimization levels, the nature of control structures can change \eg a \verb+while+ loop is replaced with a \verb+do while+ loop.

    \item \verb+puts+: Similar to \verb+printf+ the \verb+puts+ can print details about the inlined function. While the distortion of the decompiled binary is less than that of \verb+printf+, we do notice that the ordering of statements can be changed, especially for longer function bodies.
    
    \item \verb+Global array marker+: We can inject a global character pointer array (String array) in a source code file and assign it to the array for each function call. Since the array is global, the compiler will not discard it since modifying or discarding such an assignment may change the program semantics. 
    
\end{enumerate}

\begin{figure}[h!]
  \includegraphics[width=\columnwidth]{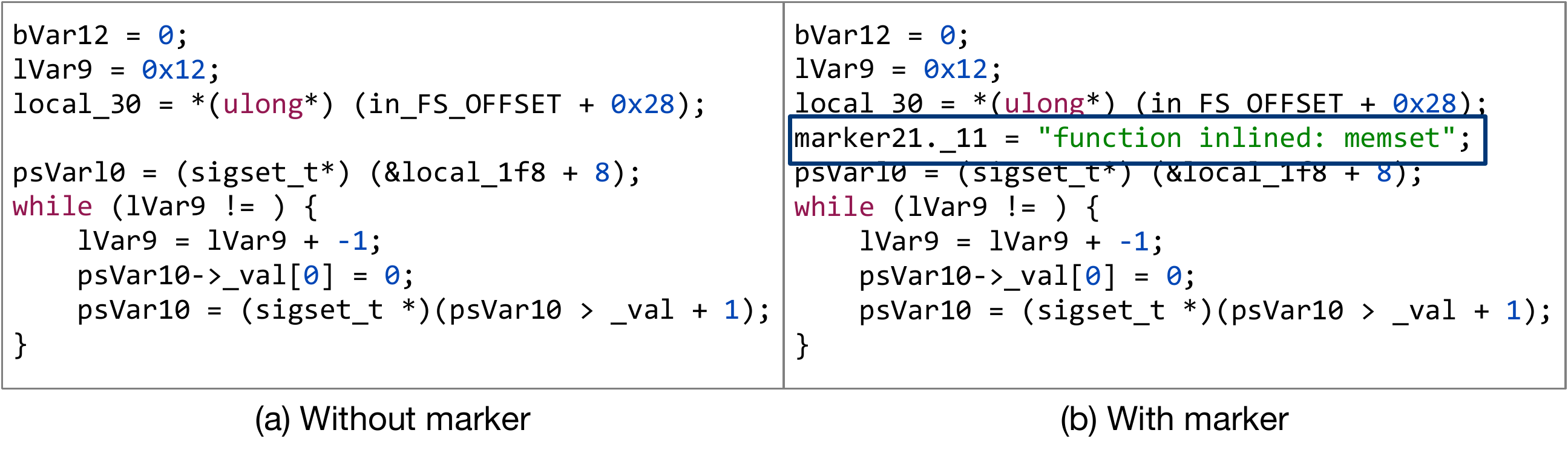}
  \caption{Marker survives -O2 optimization level without inducing any change in the code}
  \label{fig:test}
\end{figure}

Of these approaches, we chose global arrays to inject markers (\Cref{fig:test} depicts one example of an injected marker in decompiled code) in the source code. In comparison to the \verb+printf+ and \verb+puts+ approaches, the decompiled code obtained from Ghidra is not distorted. This might be due to Ghidra having a harder time in recovering library function calls in the correct position as opposed to array assignment. Furthermore, this tactic ensures that the compiler does not optimize the array access by vectorizing it, which would be the case for linear assignment. The global array we inject is a constant character pointer array of size 2000. We declare a global array and assign each marker to a different position of the array. We note that the \emph{actual value} in the array is not important for labeling; it's the assignment statement itself that constitutes the label.

In each file, we inject a uniquely identified global array, and this helps avoid a compilation conflict. This is necessary because, during compilation, the compiler merges different files (\eg header files merged into the declaring C file), which might result
inadvertently inserting multiple declarations of the same array in one file. For each function call, we assign a marker to a unique position of the array with the name of the function as seen in \Cref{fig:test}.

If we mark all function calls, we might mark some recovered by Ghidra. Since the learner does not need
to recover these, we don't mark them in the code. To remove these markers, for each function definition in the decompiled code, we compare the decompiled function definition bodies with their respective function definition in the original C code. In some cases, function calls
from the original C code that are inlined during compilation might be found by
the decompiler and indicated as such in the decompiled code. 
For those function calls that Ghidra recognizes, we remove the marker from the decompiled code; for the rest, we leave the marker as they are an indication of which function call has been inlined and where it has been inlined. 

\unpara{Identifying target functions} \label{sec:target} For this paper, we would like to design an approach that is global \ie that works on every function that has been inlined. However, for a deep learning-based approach to work, the model has to see one or more examples of an inlined function at training time, allowing it to learn an appropriate representation of each inlined function.

Using our dataset of C projects, we select a set of library functions that could be inlined in the code. We determine the most popular library function calls made by parsing all function calls from the entire dataset of 10,000 projects. To understand which function has been called in a file, we use SrcML~\cite{collard2011lightweight} to build and resolve the AST and recover the exact function binding.

After parsing all 10,000 projects, we obtain a list of the top 1,000 popularly invoked library functions that we can potentially target. From this 1,000, we filter out those that are never invoked in the 2,634 Travis-based projects, resulting in 724 potential target functions.

\subsection{Final Dataset Details}
\label{subsec:dataset}
We build and obtain binaries from 1,185 (out of 2,634) projects. Many (1,449) projects would not build. For others (726), we cannot find an exact alignment between the source code and the object files. This is because the compiler merges several source files into a single object file, thus confusing file boundaries. In such cases, it is  difficult to find the alignment between decompiled
pseudo-C, and the original source code, to allow the labeling of the pseudo-C with the requisite inlined functions. 
However, this decompiled code is still valuable, and we keep it in our dataset for (RoBERTa) pre-training purposes (as described in the next section). 

\begin{table}[h!]
\centering


\begin{subtable}[t]{\linewidth}
\centering
\caption{Cross-file train-test file distribution}
\begin{tabular}[t]{lrrr}
\hline
\multicolumn{1}{c}{\textbf{OPT-Level}} & \multicolumn{1}{c}{\textbf{Train Set}} & \multicolumn{1}{c}{\textbf{Validation Set}} & \multicolumn{1}{c}{\textbf{Test Set}} \\ \hline
O1                                     & 9                                      & 0                                           & 29                                    \\
Os                                     & 2851                                   & 88                                          & 197                                   \\
O2                                     & 2710                                   & 74                                          & 196                                   \\
O3                                     & 2591                                   & 94                                          & 144                                   \\
Of                                     & 482                                    & 14                                          & 150                                   \\ \hline
Overall                                  & 8643                                   & 270                                         & 716                                  
\end{tabular}                                
                              
\end{subtable}

\bigskip

\begin{subtable}[t]{\linewidth}
\centering
\caption{Cross-project train-test file distribution}
\begin{tabular}[t]{lrrr}
\hline
\multicolumn{1}{c}{\textbf{OPT-Level}} & \multicolumn{1}{c}{\textbf{Train Set}} & \multicolumn{1}{c}{\textbf{Validation Set}} & \multicolumn{1}{c}{\textbf{Test Set}} \\ \hline

O1        & 37        & 2              & NA       \\
Os        & 2840      & 92             & 195      \\
O2        & 2686      & 91             & 207      \\
O3        & 2517      & 96             & 215      \\
Of        & 627       & 19             & NA       \\ \hline
Overall     & 8707      & 300            & 617

\end{tabular}                                
                              
\end{subtable}


\caption{File-level distribution of the dataset used to train and test \approach\label{tab:distr_file}}
\end{table}

For the other 459 projects, we split the files into training, validation, and test sets in two different settings (file level breakdown presented in \Cref{tab:distr_file}): (1) cross-file where files from the same project can be present in the train, test or validation set and (2) cross-project where all the files from a single project are in one of the train, validation or test sets. In the cross-project setting we do not have enough projects and files for the test set for O1 and Of and thus all evaluation in this setting is done on just three settings. In the cross-file setting, our training set consists of the bodies of 391,967 function definitions spanning 8,643 files and in the cross-project setting we have 401,923 function definitions and 8,707 files. These function bodies are labeled with markers indicating any inlined functions not recovered by Ghidra and used
to construct pairs as indicated in~\Cref{fig:data_pipeline} in our dataset.

\section{Creating \approach}
\label{sec:model}

\begin{figure*}[ht!]
    \includegraphics[width=\textwidth]{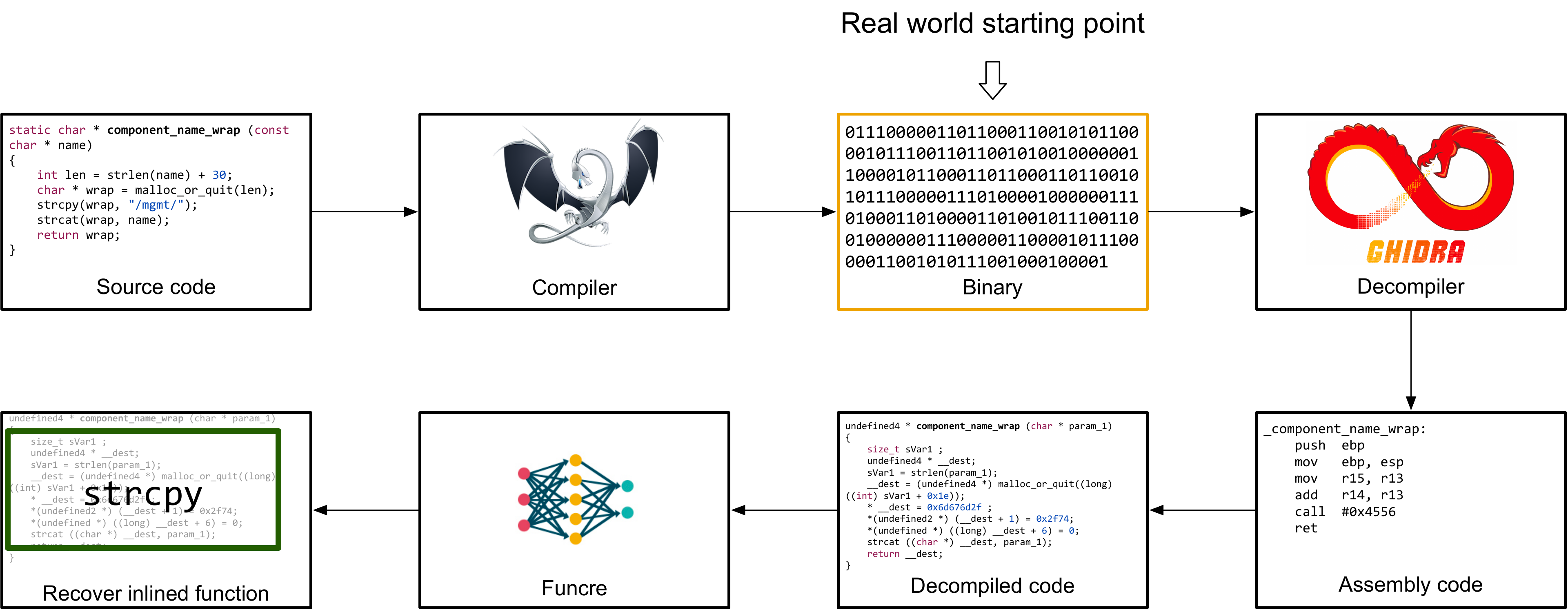}
    \caption{\label{fig:funcre_working}Working of Funcre. Funcre works on the decompiled output from Ghidra. In a real-world scenario, we start with an external binary. For training and testing purposes we create our own binaries using real-world source code obtained from GitHub.}
\end{figure*}

The expected use-case scenario of \approach is shown in \Cref{fig:funcre_working}. To get \approach working, 
we made several engineering decisions concerning the use of machine-learning techniques. 
First, we had to select a suitable
deep-learning approach. Second, we had to develop an approach to train our models. Finally, we had to design an evaluation methodology to gauge the value added by \approach.

\subsection{Model Selection}
\label{sec:model_description}

We claim that the task of recovering library function invocations from decompiled pseudo C code resembles text classification in Natural Language Processing (NLP). This intuition's essence:  function invocations, especially if inlined, can span multiple lines in decompiled pseudo-C code;  some of these lines may contain some  pattern that indicates the presence of
an invoked library function. 
We hypothesize that such patterns of pseudo-C code, reflecting the presence of library function invocations, 
 can be learned by a machine learning model given sufficient data and model capacity. In addition, our goal is to \emph{build on top} of the available tools
 that already recover atleast some function invocations, thus providing greater value to reverse engineers. 

\unpara{Potential approaches}
We consider two approaches which are the current state of the art in NLP: \emph{Transformers}, and \emph{Masked Language models with fine-tuning}. 

\smallskip

\noindent{\underline{\emph{Transformers.}} The Transformer~\cite{vaswani2017attention} model has proven to be very effective for NLP tasks. It is a sequence-to-sequence architecture (\ie it transforms a given input sequence into another sequence) consisting of an encoder and decoder. The encoder reduces an input sequence into a high dimensional vector, which is fed to the decoder, which outputs another sequence. Older sequence-to-sequence models use a RNN (Recurrent Neural Network) for the encoder and the decoder. Rather than recurrence, Transformers use a multi-head attention architecture along with feed-forward layers. Attention is a mechanism whereby for each token in an input sequence, a model chooses another token therein to ``attend to'', \emph{viz}, weight
in its decision making.  A transformer can attend to many tokens in the input sequence,  to produce an embedding of an input sequence, using ``multi-head'' attention, which improves capacity and parallelism beyond RNN  (including LSTM and GRU) approaches~\cite{vaswani2017attention}.

\noindent{\underline{\emph{Masked Language Model with fine-tuning.}}
BERT-based (Bidirectional Encoder Representations from Transformers)~\cite{devlin2018bert}  masked language models (MLM) leverage self-supervised ``pre-training". BERT learns a  representation for an input sequence.  
 A BERT model is pre-trained on large unlabeled corpora, using self-supervision, and then fine-tuned for a specific task using standard supervision (\emph{viz.,} explicitly labeled data). 
 This set-up has been shown to outperform traditional Transformer based approaches in NLP. 
For pre-training, two self-supervised tasks are used: first, it 
learns to predict masked tokens in the input sequence (typically 15\% of the tokens are masked), and second, learning
to predict the next sentence following an input sentence (NSP). This pre-trained model is then fine-tuned for a downstream supervised task such as sequence tagging. Currently, it is more common to use ROBERTA (A Robustly Optimized BERT Pretraining Approach)~\cite{liu2019roberta} to train a MLM. ROBERTA differs from BERT, with a few changes such as dynamic masking and not doing NSP, but achieves better performance. This setup of pre-training and fine-tuning achieves SOTA results for downstream SE tasks such as variable misuse detection and function-comment mismatch detection~\cite{kanade2019pre,feng2020codebert}. We reuse Huggingface's open-source implementation of ROBERTA~\cite{huggingface}.

In our setting, labeled training data (for the fine-tuning stage)
is somewhat limited, because: \begin{inparaenum}[(1)] \item we limit the task of function-invocation recovery to only a select set of library functions (see \Cref{sec:target}), \item the number of markers that we are successfully able to insert in the code and recover after the compilation - decompilation loop (see \Cref{sec:marking}) is limited, \item only projects that are compiled with a higher optimization level (O2, O3, Os, Of and in rare cases O1) can contain a function invocation that is not already recovered by Ghidra (see \Cref{sec:inline})\end{inparaenum}. At lower optimization levels, the function calls are not inlined and
are more easily found by Ghidra and Hexrays; we don't need to learn patterns for the invocations that 
are recovered already. 
As a result of these restrictions, we have a  labeled dataset that is smaller than ideal for
these powerful data-hungry transformer models. Thus, the fine-tuning approach (pre-training MLM using ROBERTA and then fine-tuning on our downstream task using the labeled dataset) is well-suited. 

For pre-training, we have available
a dataset with 1.2B tokens of  Pseudo-C files produced by Ghidra, \emph{sans any markers}.  
We omit markers here to preclude any chances of 
inadvertent ``leaking'' knowledge relevant
to the final task. Pre-training does learn robust, well-generalized representations of the statistics of decompiled
pseudo-C, which enables \approach to quickly patterns that reflect library function invocations, from a few labeled examples. 
We use the standard pre-training configuration, \emph{viz.,} ``ROBERTA base". This configuration has 12 attention layers, 768 hidden dimensions, and 12 self-attention heads in each layer resulting in a model of 125M parameters. We tokenize the code by using a Byte Level BPE (Byte Pair Encoding) Tokenizer. We limit the vocabulary size to 25,000 and keep tokens with a minimum frequency of 20. We train the MLM model on two NVIDIA Titan RTX GPUs for 2.5 epochs with a batch size of 40 sequences. This
pre-training process takes three days and achieved a final perplexity of 1.22 when predicting masked tokens. This
corresponds to a rather low cross-entropy loss of around  0.36 bits. 
This suggests that the BERT model is learning a
very powerful model of token co-occurrence statistics in the Pseudo-C
using the enormous (1.2B token) pretraining data.  For comparison, the original ROBERTA paper 
(for natural language) reported
a pre-training final perplexity as low as 3.68 (cross-entropy about  1.8 bits); the significantly
higher perplexity for natural language is consistent with prior studies~\cite{casalnuovo2019studying}. 
 
We end the pre-training once both the training and evaluation loss stops declining further. 


\smallskip

\noindent{\underline {\emph{Choosing a Context window size}}}
Before finalizing our model, we needed to address two design issues. 
\begin{inparaenum}[(1)]
\item We need to determine whether the pre-trained ROBERTA model provides any advantage over simply
training a state-of-the-art transformer model directly on task. 
\item We need to select a context window-size (in terms of the number of lines the model needs
to look at it) that can capture the signature of an in-lined method. Inlined library functions may span multiple lines. 
If this context window is too narrow, then
a pattern  capturing a library function invocation (especially if inlined)
may not fit in entirely. If it is too big, then it will compromise our
ability to locate the method more precisely. Our marking approach indicates the start of the function; however, it does not indicate how many lines it spans. We need to determine what the size of the context window (relative to the start position) must be for a model to effectively learn the representation of a function.
\end{inparaenum}

\begin{table}[]
\centering

\resizebox{\columnwidth}{!}{%
\begin{tabular}{lrrrrrr}

\hline

\multicolumn{1}{c}{\multirow{4}{*}{\textbf{Context Length}}} & \multicolumn{6}{c}{\textbf{Models}}                                                                                                                                                                                                                                                                                                                                                                                                                                                                                                                            \\
\multicolumn{1}{c}{}                                         & \multicolumn{3}{c}{\multirow{2}{*}{\textbf{Roberta-base}}}                                                                                                                                                                                                                    & \multicolumn{3}{c}{\multirow{2}{*}{\textbf{Transformer}}}                                                                                                                                                                                                                      \\
\multicolumn{1}{c}{}                                         & \multicolumn{3}{c}{}                                                                                                                                                                                                                                                          & \multicolumn{3}{c}{}                                                                                                                                                                                                                                                           \\
\multicolumn{1}{c}{}                                         & \multicolumn{1}{c}{\textbf{\begin{tabular}[c]{@{}c@{}}Top 1\\ Acc. in \%\end{tabular}}} & \multicolumn{1}{c}{\textbf{\begin{tabular}[c]{@{}c@{}}Top 5\\ Acc. in \%\end{tabular}}} & \multicolumn{1}{c}{\textbf{\begin{tabular}[c]{@{}c@{}}Top 10 \\ Acc. in \%\end{tabular}}} & \multicolumn{1}{c}{\textbf{\begin{tabular}[c]{@{}c@{}}Top 1\\ Acc. in \%\end{tabular}}} & \multicolumn{1}{c}{\textbf{\begin{tabular}[c]{@{}c@{}}Top 5 \\ Acc. in \%\end{tabular}}} & \multicolumn{1}{c}{\textbf{\begin{tabular}[c]{@{}c@{}}Top 10 \\ Acc. in \%\end{tabular}}} \\ \hline
$\pm3$                                                          & 75.73                                                                                   & 88.33                                                                                   & 91.44                                                                                     & 64.93                                                                                   & 84.56                                                                                    & 89.28                                                                                     \\
$\pm5$                                                           & 80.38                                                                                   & 91.64                                                                                   & 95.02                                                                                     & 71.71                                                                                   & 90.12                                                                                    & 94.14                                                                                     \\
$+10$                                                          & 78.28                                                                                   & 91.25                                                                                   & 93.45                                                                                     & 71.92                                                                                   & 89.20                                                                                    & 93.38                                                                                     \\
$-10$                                                          & 56.43                                                                                   & 76.64                                                                                   & 83.17                                                                                     & 49.00                                                                                   & 73.17                                                                                    & 81.38                                                                                     \\
$\pm10$                                                          & \textbf{80.41}                                                                          & \textbf{92.56}                                                                          & \textbf{95.21}                                                                            & 69.37                                                                                   & 89.41                                                                                    & 93.13                                                                                   
\end{tabular}
}
\caption{Performance of RoBERTa and Transformer models on development set at different context size\label{tab:withloc}}
\end{table}

\unpara{Finalizing our Design}
We evaluated our design choices on our validation set, using a straw-man task.  
We train a vanilla Transformer model end-to-end on the labeled training dataset. For the ROBERTA-based model, we reuse the pre-trained MLM mentioned earlier and then fine-tune it on our labeled dataset. 
We train the models by feeding them input sequences where the  function invocation marker is in the middle surrounded by a context window (in terms of the number of lines) of varying sizes. This form of training does not parallel a real-world setting where the location of function  invocation is typically unknown; however, to choose a model
architecture, this approach is reasonable. 

In~\Cref{tab:withloc} we see the performance of both the ROBERTA model and the Transformer model on five different context window sizes. We observe that the window size $\pm 10$ works best in the case of ROBERTA. We also notice that the context window sizes $+10$ and $-10$ do not work as well as a context window that spans both sides of a marker. This implies that the model requires both the preceding and succeeding lines of code to learn the representation of a function invocation. 
Both ROBERTA and the Transformer model learn a fairly good representation of the invocations. With the Top 1 accuracy for ROBERTA reaching 80\% and the Top 5 accuracy being 92.5\% (both in the case of context window size $\pm 10$). Furthermore, we see that in \Cref{tab:withloc} the ROBERTA model outperforms the Transformer model in every setting. This implies that by pre-training a MLM and then fine-tuning it on the task, we can achieve better performance. 

\emph{We, therefore, chose the  MLM architecture, and $\pm 10$ context window size, for our more
realistic training/evaluation regime}.

\subsection{Final Training \& Evaluation}
\label{sec:block_eval}
Using the pre-trained MLM, now we fine-tune the MLM using the labeled dataset to create \approach.
In the earlier straw-man approach, which was used model selection, 
we assumed that the model would know the location of the library function invocation. 
In a real-world setting,  the model 
must recover the invocation from the entirety of the decompiled code 
for a function definition, without knowledge of the specific location. Because of optimizations such as inlining and code movement, it's often not possible to know exactly where in the pseudo-C the function invocation should be located; we therefore relax the problem to locating the \emph{calling function body} in which the invocation occurs. A correct recovery for us would therefore amount to recovering the correct invoked method, and the correct function body in which that method is invoked. Our recovery works by scanning a window over the a function body and looking for invocations within each window. 

\unpara{Scanning the Window} Based on the indication from the straw-man evaluation above, 
we employ a context window of $\pm 10$ size to both train and test our model. 
For function bodies that are over 20 lines long, we slide a context window forwards, one line at a time. 
In both training and test, each sliding window is labeled with the marker that occurs in that window. When
there are multiple markers in a window, we simplify the labeling for the block by marking it with the first marker in lexical order.
This line-by-line scanning does present a problem. Consider an inlined library function invocation (say \verb+atoi+)  that occurs at line 8
of a 30-line function. The corresponding marker will occur around line 8 in the first 20-line scanning window (starting
at line 1 of the function) and repeat eight or more times as the 20-line scanning window moves forward, a line at a time. The
learner may find an invoked function's signature in even more than eight successive windows.
Consequently, we adopt a sequential filtering heuristic to coalesce these repeating 
sequential labels, as described next.

\unpara{Coalescing Sequential Labels}
We use a simple, noise-tolerant run-length encoding heuristic to coalesce sequential blocks. This heuristic was
tuned on the validation set without examining the test set.

\begin{enumerate}
\item 
Given a sequence of predicted labels in long function, we remove sequentially inconsistent predictions, 
\emph{viz}, labels that disagree with the five preceding \emph{and} succeeding labels. This works well since in-lining is rare in shorter functions. 
\item We use run-length encoding on the sequence. Run-lengths are incremented leniently; 
our heuristic will treat up to 3 successive unlabeled windows as being the same as preceding label and succeeding label (if both preceding label and succeeding label are same). 
Thus if $a, b, c$ are method labels and
$x$ is a ``no method'' label, the label sequence $aaxxaabbbbbxcxxxxcdc$ is encoded as $a^6b^5c^1c^3$. 
The second two $x$ after the first two $a$ are treated as $a$, so we get a total run length of 6 $a$; 
after the 5 $b$, the $c$ is accepted, although it is inconsistent with the preceding $b$ because it agrees
with a following $c$; the $d$ within the run of 2 $c$ at the end is erased for being inconsistent. 

\item Next, we only retain function labels with a run-length of at least four as a true label. The above example
then collapse to just $a^6b^5$. 
Finally, we divide the run-length by 20 and take the ceiling. This leaves us with just two labels, $a$ and a following $b$. 
We collect the markers from the Ghidra output and compare the result. Finally, we add the result to the result achieved by Ghidra. 
\end{enumerate}

We note again that this heuristic was tuned \emph{{\bf exclusively} on the validation set}, to 
scrupulously avoid overfitting to the test set.

\unpara{Data Imbalance} After dividing both the training and validation sets into windows containing 20 lines of code shifted by one line, we observe that our dataset is imbalanced:  the unlabeled windows dominate.  Since we have a pre-trained ROBERTA model that has  learned the statistics of the unlabeled window, we re-balance the data 
by discarding some 65\%
of these blocks with no label. 
Even so, the no-label windows are about 80\% of our training and development set.
For the fine-tuning stage, we employ the same tokenizer that was used for pre-training.
We fine-tune our model over three epochs on six NVIDIA Titan RTX GPUs, taking a total of three hours.

\section{Empirical results}
Our goal is to improve the performance of Ghidra in recovering inlined library functions. 
Ghidra already recovers some library functions; the combination
of Ghidra with our model \emph{should} improve performance. We begin with our evaluation metric
and then dive into our results. 

\unpara{Evaluation Metric}


We remind the reader (as described in \Cref{sec:block_eval} ) a correct recovery for us is a library function invocation, together with the function
body in which this invocation occurs. 
Thus \emph{a single test instance is a specific library function invocation, together with the containing function; this is our target for recovery}.  
This task is performed by scanning candidate function bodies in 20-line blocks; we explain above 
\Cref{sec:block_eval}  how the model decides if and what function invocations occur in these blocks. In the following we explain our evaluation criteria as it applies to the problem of \emph{recovering library function invocations within function bodies}, \emph{viz.} 
how we decide if a test instance results in a TP, FP, TN, FN, \emph{etc}
\begin{itemize}
    \item \emph{Model predicts empty (\ie no invoked function) and true label is empty: this function body contains no library function invocations} we mark this as a true negative (TN). These are of limited value to the RE and extremely numerous!  So we ignore these in our calculations (note that we do not report ``accuracy'', and do not count these in our precision calculation). 
      \item \emph{Model predicts `Func1' and true label is `Func1':}  we count it as a true positive (TP). These are helpful to the RE. 
    \item \emph{Model predicts label `Func1', and true label is empty:} we count a false positive (FP). In this case, our model is confused, and finds a library
    function invocation in the body of another function, where there isn't any such invocation. These create needless work for the RE. 
     \item \emph{Model predicts empty, and true label is `Func1'} we count a false negative (FN). Our model failed to recover a library function invocation that did occur within a function body. 
     These cases fail to provide useful information to the RE. 
    \item \emph{Model predicts label `Func1' and true  label `Func2:}  we score this as a false negative, FN, (for missing `Func2'), and also a FP (for predicting `Func1').  Our model not only failed to recover a function invocation, it also reported incorrectly that a different function was used than the actual one! These cases fail to report important information, \emph{and} create needless work, and so is doubly penalized. 
\end{itemize}

As can be seen from the above, TP+FN is the count of actual function invocations that \emph{could} be recovered. 
Based on these counts for FP, FN, TP, and TN, we calculate the precision, recall, and F-score that our model achieves on the test set. 
Note again that we ignore the correctly labeled empties despite this being an instance of our model performing well (this lowers our precision from $\sim$0.90 to $\sim$0.60) since it is of limited value to the RE. All the evaluations are given below use these criteria.


\subsection{RQ1: Effectiveness of \approach}

\begin{table}[h!]
\centering


\begin{subtable}[t]{\linewidth}
\caption{Cross-file train-test split\label{tab:with_loc_file}}
\begin{tabular}[t]{lrrrrrr}
\hline
\multicolumn{1}{c}{\textbf{OPT-Level}} & \multicolumn{1}{c}{\textbf{TP}} & \multicolumn{1}{c}{\textbf{FP}} & \multicolumn{1}{c}{\textbf{FN}} & \multicolumn{1}{c}{\textbf{Prec.}} & \multicolumn{1}{c}{\textbf{Recall}} & \multicolumn{1}{c}{\textbf{F-score}} \\ \hline
O1                                     & 135                             & 86                              & 130                             & 0.61                               & 0.51                                & 0.56                                 \\ 
Os                                     & 752                             & 366                             & 867                             & 0.67                               & 0.46                                & 0.55                                 \\
O2                                     & 647                             & 403                             & 760                             & 0.62                               & 0.46                                & 0.53                                 \\
O3                                     & 736                             & 437                             & 955                             & 0.63                               & 0.44                                & 0.51                                 \\
Of                                     & 898                             & 429                             & 998                             & 0.67                               & 0.47                                & 0.56 \\ \hline
Overall                                & 3168                            & 1721                            & 3710                            & 0.64                               & 0.46                                & 0.54                                 
                              
\end{tabular}
\end{subtable}

\bigskip

\begin{subtable}[t]{\linewidth}
\caption{Cross-project train-test split\label{tab:with_loc_project}}
\begin{tabular}[t]{lrrrrrr}
\hline
\multicolumn{1}{c}{\textbf{OPT-Level}} & \multicolumn{1}{c}{\textbf{TP}} & \multicolumn{1}{c}{\textbf{FP}} & \multicolumn{1}{c}{\textbf{FN}} & \multicolumn{1}{c}{\textbf{Prec.}} & \multicolumn{1}{c}{\textbf{Recall}} & \multicolumn{1}{c}{\textbf{F-score}} \\ \hline
O2        & 1004 & 417  & 923  & 0.70  & 0.52   & 0.60    \\
O3        & 743  & 1192 & 1222 & 0.38  & 0.38   & 0.38    \\
Os        & 927  & 641  & 919  & 0.59  & 0.50   & 0.52    \\

\hline

Overrall  & 2674 & 2250 & 3064 & 0.54  & 0.46   & 0.50                          
                              
\end{tabular}
\end{subtable}


\caption{Performance of \approach at various optimization levels}
\end{table}

Now, we start with the results for \approach  on the task of recovering library function invocations \emph{not} recovered by Ghidra (we evaluate
the recovery of \emph{all} function invocations later). 
Our evaluation is based on two different dataset splits as elucidated in \Cref{subsec:dataset}. \Cref{tab:with_loc_file} presents our model's performance on the test set that has been split at file level. Overall test instances, \approach achieves a precision of 0.64 and a recall of 0.46. We have more FNs than FPs, suggesting that the model errs on the side of caution \ie instead of inaccurately predicting the presence of a function, the model predicts that there is no function in place. 

In \Cref{tab:with_loc_project} we present our results on a test that has been split at project level. With this test split we only have sufficient training and test data for three optimization levels - O2, O3, Os as for O1 and Of the data originates from a few projects thus rendering a project level split impossible. We observe that with this cross project split, the precision and recall at O2 increases in comparison with the cross-file split. We see that for Os we lose some precision but improve the recall and overall the F-score is slightly lower. However, for O3 we see a degradation in both precision and recall. 

\begin{figure*}[h!]
\centering
    \includegraphics[width=\textwidth]{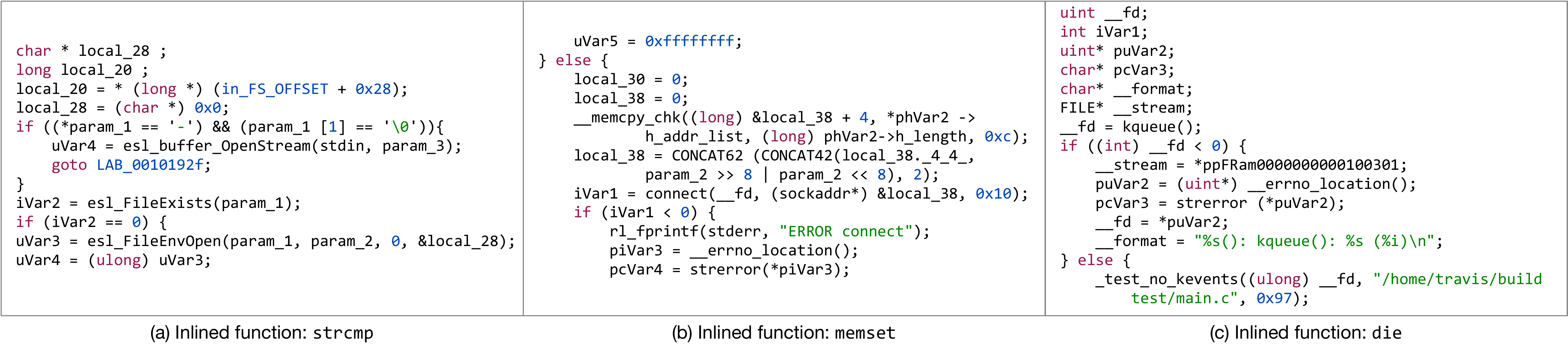}
    \caption{Example code snippet from decompiled code containing an inlined library function\label{fig:sample}}
\end{figure*}
Since the data in our cross-file dataset is so imbalanced, we check if our model performs better than a random model or coin toss.
We made several runs of a simulated model that uses only prior probability to predict invoked function. 
Not surprisingly,  \approach vastly outperforms a simulated model that guesses labels just based
on priors: the f-score never rises above 0.003 despite hundreds of simulated runs.

\Cref{fig:sample} shows three code snippets from our test set which contain the functions \verb+memset+, \verb+strcmp+, and \verb+die+ respectively. Despite the lack of obvious signs in the decompiled code of these invoked functions, our model can identify the functions correctly. This suggests that our model is learning a useful representation of the decompiled code and can recover (even inlined) function calls. Table~\ref{list_of_funcs} presents a sample list of library functions recovered by Funcre. Several of these represent
vulnerabilities, and/or malicious behavior; in general, labeling unidentified 
library function calls correctly,  in decompiled code, represents useful information that 
is \emph{simply not} otherwise available to the reverse engineer.

\begin{table}[h!]
\ttfamily
\centering
\resizebox{\columnwidth}{!}{%
\renewcommand{\arraystretch}{1.2}
\begin{tabular}{|l|}
\hline

\begin{tabular}[c]{@{}l@{}}memset, fprintf, check, setjmp, match, snprintf, free,\\fopen, wifexited, closesocket, xmalloc, htons, calloc,\\testnext, malloc, open, localtime, wifsignaled, impossible,\\fail, unlock, xstrdup, pixgetdata, verbose, validate, typeof,\\getpid, strcasecmp, warnx, waitforsingleobject, getgid,\\system, entercriticalsection, createevent, setsockopt, raise\\crc32, leavecriticalsection, perror, chmod, report

\end{tabular} \\ \hline
\end{tabular}
}
\caption{Functions recovered by \approach. \label{list_of_funcs}}
\end{table}


\takeaway{1}{Overall, the code representation learned by \approach is powerful enough to recover 46\% of library function calls, and errs on the side of caution (more FN than FP)}

\subsection{RQ2: Effect of Optimization Level}
We look at the effect of optimization level on \approach in only the cross-file setting as the cross-project setting has not been evaluated on two optimization levels.
In \Cref{tab:with_loc_file} we examine model performance at varying optimization levels: higher optimization
levels make the task harder.   
For the O1 level, we have reduced training data due to the low rate of function inlining that occurs; thus we see the poorest performance. Likewise, we have reduced training data for Of; however, we see that model's precision is higher in this case. This despite Of being the most complex optimization;  
we hypothesize that is because of inductive transference from the O3, O2, and Os, classes, 
where many similar optimizations may occur. 

For Os, O2, and O3, the F-score ranges between 0.55 and 0.51. Additionally, in all three cases, the precision is higher than 0.60. As the optimization level increases, the precision drops slightly. However, the recall remains almost constant. This relatively stable performance suggests that the model can deal with the challenges posed by higher optimizations, where the decompiler typically struggles to create a helpful representation.

\takeaway{2}{The performance of \approach does not deteriorate significantly with the increasing complexity of compiler optimization.}

\subsection{RQ3: Impact of the popularity of methods}

\begin{figure}[h]
    \begin{center}
        \includegraphics[width=0.94\columnwidth]{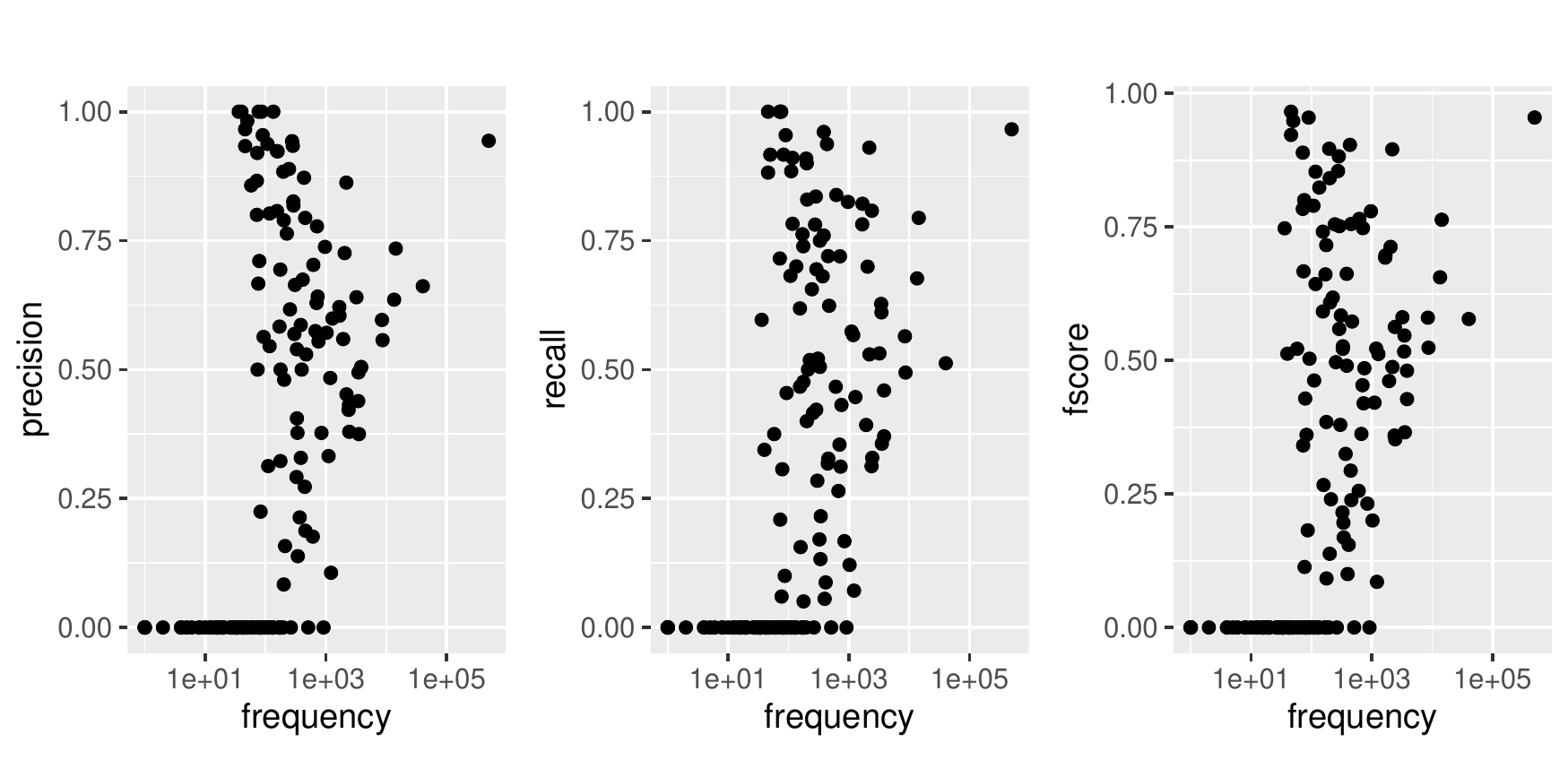}
    \end{center}
    \caption{Does training set frequency affect model performance?\label{fig:freq}}
\end{figure}

How does performance vary with training sample frequency? \Cref{fig:freq} plots the method frequency in the training set against precision, recall, and F-score in the cross-file setting (we omit the cross project setting here due to the relative lack of diversity of the test set). We can see that for methods that occur less than roughly 50 times in the training dataset, performance is generally quite low. These methods include \verb+strtoi+ (frequency is 1), \verb+vprintf+ (frequency is 20) and \verb+rand+ (frequency is 30). 
At intermediate frequencies, between 50 and 1500, performance is quite variable. 
There are some popular methods such as \verb+offsetof+ (frequency is 917) and \verb+max+ (frequency is 1,220) for whom the F-score remains quite low, near 0. However \approach can perform well on functions such as \verb+lseek+ (frequency is 90) and \verb+strndup+ (frequency is 72) where the F-score is higher than 0.8. We conjecture that performance depends
on other factors, \eg how varied the invocation (or inlined) code looks for each function.

At much higher frequencies, performance more reliably improves;  methods such as \verb+sscanf+ (frequency is 2,186), \verb+printf+ (frequency is 14,437) and \verb+assert+ (frequency is 40,520) all show good performance. 
\approach is also able to perform well on rares functions such as \verb+lseek+ (frequency is 90) and \verb+strndup+ (frequency is 72) where the F-score is higher than 0.8. 

The overall Pearson correlation values of the frequency with precision is 0.14,  with recall  0.16, and with F-score 0.17.


\takeaway{3}{The performance of \approach has a weak correlation with call frequency. 
The weak correlation arises from 3 distinct regions of performance: consistently poor performance at low
frequencies, very variable in mid-ranges, and more reliably higher at higher frequencies.}

\subsection{RQ4: Comparison to existing tools}
\label{sec:compare}
\begin{table}[h!]

\centering

\begin{subtable}[t]{\linewidth}
\caption{Cross-file train-test split\label{tab:ghidra-hex-file}}
\resizebox{\columnwidth}{!}{%
\renewcommand{\arraystretch}{1.2}

\begin{tabular}[t]{lrrlrrrrrrr}
\hline

\multicolumn{1}{c}{\textbf{\begin{tabular}[c]{@{}c@{}}OPT.\\ level\end{tabular}}} & \multicolumn{1}{c}{\textbf{\begin{tabular}[c]{@{}c@{}}File\\ Count\end{tabular}}} & \multicolumn{1}{c}{\textbf{\begin{tabular}[c]{@{}c@{}}Total  \\ Functions\end{tabular}}} & \multicolumn{1}{c}{\textbf{Tool}} & \multicolumn{1}{c}{\textbf{TP}} & \multicolumn{1}{c}{\textbf{FP}} & \multicolumn{1}{c}{\textbf{FN}} & \multicolumn{1}{c}{\textbf{Prec.}} & \multicolumn{1}{c}{\textbf{Recall}} & \multicolumn{1}{c}{\textbf{F-score}} & \multicolumn{1}{c}{\textbf{\begin{tabular}[c]{@{}c@{}}Unique\\  Function \\ Recovered\end{tabular}}} \\ \hline

\multirow{3}{*}{O1}                                                               & \multirow{3}{*}{29}                                                               & \multirow{3}{*}{867}                                                                     & Hex-Rays                          & 456                             & 103                             & 407                             & 0.81                               & 0.53                                & 0.64                                 & 50                                                                                                   \\ 
                                                                                  &                                                                                   &                                                                                          & Ghidra                            & 460                             & 67                              & 404                             & \textbf{0.87}                      & 0.53                                & 0.66                                 & 51                                                                                                   \\ 
                                                                                  &                                                                                   &                                                                                          & Ghidra+\approach                         & 595                             & 153                             & 400                             & 0.80                               & \textbf{0.59}                       & \textbf{0.68}                        & \textbf{54}                                                                                          \\ \hline

\multirow{3}{*}{Os}                                                               & \multirow{3}{*}{197}                                                              & \multirow{3}{*}{2932}                                                                    & Hex-Rays                          & 1796                            & 1819                            & 2665                            & 0.50                               & 0.40                                & 0.44                                 & 100                                                                                                  \\
                                                                                  &                                                                                   &                                                                                          & Ghidra                            & 1632                            & 1399                            & 2983                            & 0.54                               & 0.36                                & 0.43                                 & 89                                                                                                   \\
                                                                                  &                                                                                   &                                                                                          & Ghidra+\approach                         & 2384                            & 1765                            & 3144                            & \textbf{0.57}                      & \textbf{0.43}                       & \textbf{0.49}                        & \textbf{108}                                                                                         \\ \hline

\multirow{3}{*}{O2}                                                               & \multirow{3}{*}{196}                                                              & \multirow{3}{*}{1686}                                                                    & Hex-Rays                          & 2234                            & 1325                            & 2764                            & 0.62                               & 0.44                                & 0.52                                 & 131                                                                                                  \\
                                                                                  &                                                                                   &                                                                                          & Ghidra                            & 2012                            & 934                             & 3129                            & \textbf{0.68}                      & 0.39                                & 0.50                                 & 126                                                                                                  \\
                                                                                  &                                                                                   &                                                                                          & Ghidra+\approach                         & 2659                            & 1337                            & 3256                            & 0.67                               & \textbf{0.45}                       & \textbf{0.54}                        & \textbf{160}                                                                                         \\ \hline

\multirow{3}{*}{O3}                                                               & \multirow{3}{*}{144}                                                              & \multirow{3}{*}{1747}                                                                    & Hex-Rays                          & 2499                            & 3330                            & 3279                            & 0.43                               & 0.43                                & 0.43                                 & 104                                                                                                  \\
                                                                                  &                                                                                   &                                                                                          & Ghidra                            & 2334                            & 2191                            & 3699                            & 0.52                               & 0.39                                & 0.44                                 & 98                                                                                                   \\
                                                                                  &                                                                                   &                                                                                          & Ghidra+\approach                         & 3070                            & 2628                            & 3950                            & \textbf{0.53}                      & \textbf{0.43}                       & \textbf{0.48}                        & \textbf{126}                                                                                         \\ \hline

\multirow{3}{*}{Of}                                                               & \multirow{3}{*}{150}                                                              & \multirow{3}{*}{1579}                                                                    & Hex-Rays                          & 1111                            & 489                             & 2840                            & \textbf{0.69}                               & 0.28                                & 0.40                                 & 78                                                                                                   \\
                                                                                  &                                                                                   &                                                                                          & Ghidra                            & 893                             & 656                             & 3282                            & 0.58                               & 0.21                                & 0.31                                 & 75                                                                                                   \\
                                                                                  &                                                                                   &                                                                                          & Ghidra+\approach                         & 1791                            & 1085                            & 3411                            & 0.62                      & \textbf{0.34}                       & \textbf{0.44}                        & \textbf{89}                                                                                         
\end{tabular}

}
\end{subtable}

\bigskip

\begin{subtable}[t]{\linewidth}
\caption{Cross-project train-test split\label{tab:ghidra-hex-project}}
\resizebox{\columnwidth}{!}{%
\renewcommand{\arraystretch}{1.2}

\begin{tabular}[t]{lrrlrrrrrrr}
\hline

\multicolumn{1}{c}{\textbf{\begin{tabular}[c]{@{}c@{}}OPT.\\ level\end{tabular}}} & \multicolumn{1}{c}{\textbf{\begin{tabular}[c]{@{}c@{}}File\\ Count\end{tabular}}} & \multicolumn{1}{c}{\textbf{\begin{tabular}[c]{@{}c@{}}Total  \\ Functions\end{tabular}}} & \multicolumn{1}{c}{\textbf{Tool}} & \multicolumn{1}{c}{\textbf{TP}} & \multicolumn{1}{c}{\textbf{FP}} & \multicolumn{1}{c}{\textbf{FN}} & \multicolumn{1}{c}{\textbf{Prec.}} & \multicolumn{1}{c}{\textbf{Recall}} & \multicolumn{1}{c}{\textbf{F-score}} & \multicolumn{1}{c}{\textbf{\begin{tabular}[c]{@{}c@{}}Unique\\  Function \\ Recovered\end{tabular}}} \\ \hline

\multirow{3}{*}{O2} & \multirow{3}{*}{207} & \multirow{3}{*}{1250} & IDA       & 2028 & 2028 & 3917 & 0.34  & 0.27   & 0.30    & 110             \\
                    &                      &                       & Ghidra    & 1864 & 1593 & 3664 & 0.53  & 0.33   & 0.41    & 105             \\
                    &                      &                       & Ghidra+Us & 2868 & 2010 & 3633 & \textbf{0.58}  & \textbf{0.44}   & \textbf{0.50 }   & \textbf{123}             \\ \hline
\multirow{3}{*}{O3} & \multirow{3}{*}{215} & \multirow{3}{*}{2496} & IDA       & 1784 & 5031 & 5323 & 0.26  & 0.25   & 0.26    & 115             \\
                    &                      &                       & Ghidra    & 1623 & 2030 & 3771 & \textbf{0.44}  & 0.30   & 0.36    & 110             \\
                    &                      &                       & Ghidra+Us & 2366 & 3222 & 4346 & 0.42  & 0.35   & \textbf{0.39}    & \textbf{134}             \\ \hline
\multirow{3}{*}{Os} & \multirow{3}{*}{195} & \multirow{3}{*}{2687} & IDA       & 2498 & 3967 & 5362 & 0.39  & 0.32   & 0.35    & 102             \\
                    &                      &                       & Ghidra    & 2363 & 1280 & 3115 & \textbf{0.64}  & 0.43   & 0.52    & 98              \\
                    &                      &                       & Ghidra+Us & 3290 & 1921 & 3242 & 0.63  & 0.50   & \textbf{0.56}    & \textbf{113}  
\end{tabular}

}
\end{subtable}

\caption{Comparison of Ghidra + \approach, Ghidra and Hexrays}
\end{table}

Next, we compare our overall performance to the state-of-the-art, Ghidra, and Hex-rays.
In this evaluation, we consider recovery of \emph{all} function invocations. How well do the available
tools identify function invocations, whether inlined or not? 
Since \approach works on top of Ghidra, we augment Ghidra's results with ours to measure if (and how much)
 the function recovery of Ghidra is improved. 

To assess the TP and FP rates for function recovery with Ghidra and Hex-rays, we compare the decompiled code against the original source code to see how many functions are recovered correctly.
We run this analysis on our test set in both the cross-file and cross-project setting. 
Compared to \Cref{tab:with_loc_file}, we are missing the results for Hex-rays on 14 files in our test set, 
because the version of Hex-rays at our disposal cannot process 32-bit binaries\footnote{A commercial license for the 32 bit version of Hexrays is available for additional purchase}, so we just omitted them from this comparison, to be
generous to Hex-rays.  

In \Cref{tab:ghidra-hex-file} we see that Ghidra + \approach has the best f-score for all optimizations (precision declines somewhat for O1 and Of, and marginally for O2). Our tool does not degrade the recall or F-score for Ghidra; instead, it enhances it enough to outperform  Hex-rays. We also see that Ghidra + \approach recovers the most library function calls, and in all cases, also the most \emph{unique} functions:  \eg at the O2 level, Ghidra + \approach recovers 29 more unique functions than Hex-rays, while achieving also getting the highest F-score. 

All outputs of \approach, Hex-ray, and Ghidra are multisets. If one {\small\tt EnterCriticalSection} and 3 {\small\tt sprintf} are actually inlined, Ghidra may recover partially (e.g, one {\small\tt sprintf} is recovered), and we combine those with our output. To combine outputs, we take a multiset union (which could boost both TP and FP, and reduce FN; note that both function name and count matter to measure Recall/Precision/F1). \Cref{tab:with_loc_file} reports on JUST the MARKED functions (we evaluate on the recovery of 2 sprintf and 1 EnterCriticalSection) recovered per optimization level by Funcre ALONE.

We repeat the same analysis in our cross project setting as well and present the results in \Cref{tab:ghidra-hex-project}. We see that in contrast to the cross file setting, the precision and recall is down in all three scenarios across all three optimization levels. Despite this downturn, Ghidra + \approach outperforms plain Ghidra and Hex Rays in terms of F-score in all the cases. We do notice that the precision and recall for O2 and O3 are lower than in the cross file setting, however, for Os there is an increase. Overall, in the cross project setting we see a drop in performance in comparison to the cross file setting, but even in this setting \approach shows that it outperforms the competition.

We find it noteworthy that the Hex-ray's FP count is this high given that the Hex-rays developers explicitly designed their FLIRT signature system to be cautious and never introduced a false positive (see \Cref{sec:inline}).

To investigate further, we examine a random subset of 10 function definitions containing one or more FP library calls for each of the three tools. We observe that in the majority (seven for Hex-rays, seven for Ghidra, and five out of ten for Ghidra + \approach) of the cases, the FP are correctly marked as FP \ie the tool incorrectly recovers the wrong function based on a comparison with the original source. In the remaining cases, we find that the function call is transitively inlined from a function definition or macro from another file and due to the limitation in our detection strategy (these cases are near impossible to detect due to the absence of a system-level call graph) are marked as FP.

Out of 724 popularly used library functions present in the projects we target , only 365 are inlined depending on optimization level (O1 has minimal inlining). Only 168 occur in our test set, and Funcre recovers 93. Significantly, we improve 19\% over Ghidra and 10\% over Hex-rays; for the challenging case of O2, we improve by 22\% over Hex-rays (second highest in \Cref{tab:ghidra-hex-file}). Note that we correctly recover more inlined-functions; our recall improves over Hex-rays e.g., for Of by over 20\%, finding 680 more instances of inlined-functions. While improving recall, Funcre leaves precision about the same or improved (\Cref{tab:ghidra-hex-file}). Our FP rate \emph{is not higher} than current tooling.


\takeaway{4}{\approach enhances the performance of Ghidra to the extent that it outperforms Hex-rays.}

\section{Threats to validity}

\noindent\textbf{Generalizability.}
We collect code and build binaries from GitHub projects;  these may not yield binaries that are typically reverse-engineered. Furthermore, we only target binaries built against three versions of Linux on an x86 architecture; performance on other platforms may vary. 


\noindent\textbf{Internal validity.}
We mark predictions as true only when exactly matched with the expected label. However, in some cases, the decompilers recover functions such as \verb+printf_chk+ and \verb+assert_fail+ rather than \verb+printf+ and \verb+assert+. One could argue that the prediction is correct in such cases, but we still mark it as incorrect. This impacts measured scores equally for all tools (\approach, Ghidra, and Hex-rays). This is a non-trivial issue with no easy resolution: \emph{e.g.,} not requiring an exact match also risks biases \& errors.

Certain false positives in the function recovery for all three tools also originate from the fact that function definitions or macros from the same project might have been inlined into the function that we analyze. Inlined calls transitively inlined in the function under consideration might be recovered by all three tools; however, we have no way of knowing whether the recovery is a true positive. We do look at this transitive inlining for one step \ie for the first function declaration inside the same file that is inlined. However, we do not construct a system-level call graph which might adversely impact the false positive rate reported for all three tools.

\noindent\textbf{Practical applicability.}
When both Ghidra and Hex-rays recover a function call, they can place it in the function definition body as an actual function call and not a collection of statements. \approach can only recover the list of inlined functions per function declaration body. However, as seen in \Cref{sec:compare} we observe that both Ghidra and Hex-rays have false positives when it comes to function recovery, furthermore, the evaluation strategy employed in this work does not know whether the location where the function is recovered is correct or whether the parameters that are passed to the function call are correct. We do recover some functions that current tools cannot; still, marking the exact position of the inlined is function is much harder because the compilation-decompilation loop can move the location of a marker. Knowing exactly which lines correspond to an inlined function is also non-trivial without a more advanced representation of the code.


\section{Contributions}
We have described an approach to improve inlined library function recovery in comparison with state-of-the-art tools (Ghidra and Hex-rays IDA Pro). Our main contributions are: 

\begin{enumerate}
    \item We created a technique to build C-based projects on a large scale. Using this pipeline, we build and release the Docker containers for 1,185 C projects. We also created an annotated set of real projects (first of its kind), which indicates functions inlined by compilers. Our data \& tooling will be released.
    \item We show that MLM pre-training, on 1.2 billion tokens of decompiled code improves task-performance for inlined library function recovery. We will release the trained MLM for reuse in other RE tasks, such as name recovery. 
    \item We improve upon Ghidra and Hex-rays on library function recovery, both in terms of f-score and unique functions recovered. This suggests that modern machine-learning methods have value for this task. 
    \item There has been less attention in prior research work (on binary analysis) towards highly optimized binaries. Our work considers all optimization settings, including the highest (Of). This suggests that our research has greater relevance in broader settings than prior work.
\end{enumerate}

\ifCLASSOPTIONcompsoc
  \section*{Acknowledgments}
\else
  \section*{Acknowledgment}
\fi

We gratefully acknowledge support from NSF CISE (SHF LARGE) Grant No. 1414172, and from
Sandia National Laboratories. Toufique Ahmed is supported by a Dean's Distinguished Graduate Fellowship. 
This paper was much improved as a result of reviewers' comments, for which we are thankful.

\ifCLASSOPTIONcaptionsoff
  \newpage
\fi



\bibliographystyle{IEEEtran}
\bibliography{decom-bib.bib}
%



%

\begin{IEEEbiography}[{\includegraphics[width=1in,height=1.25in,clip,keepaspectratio]{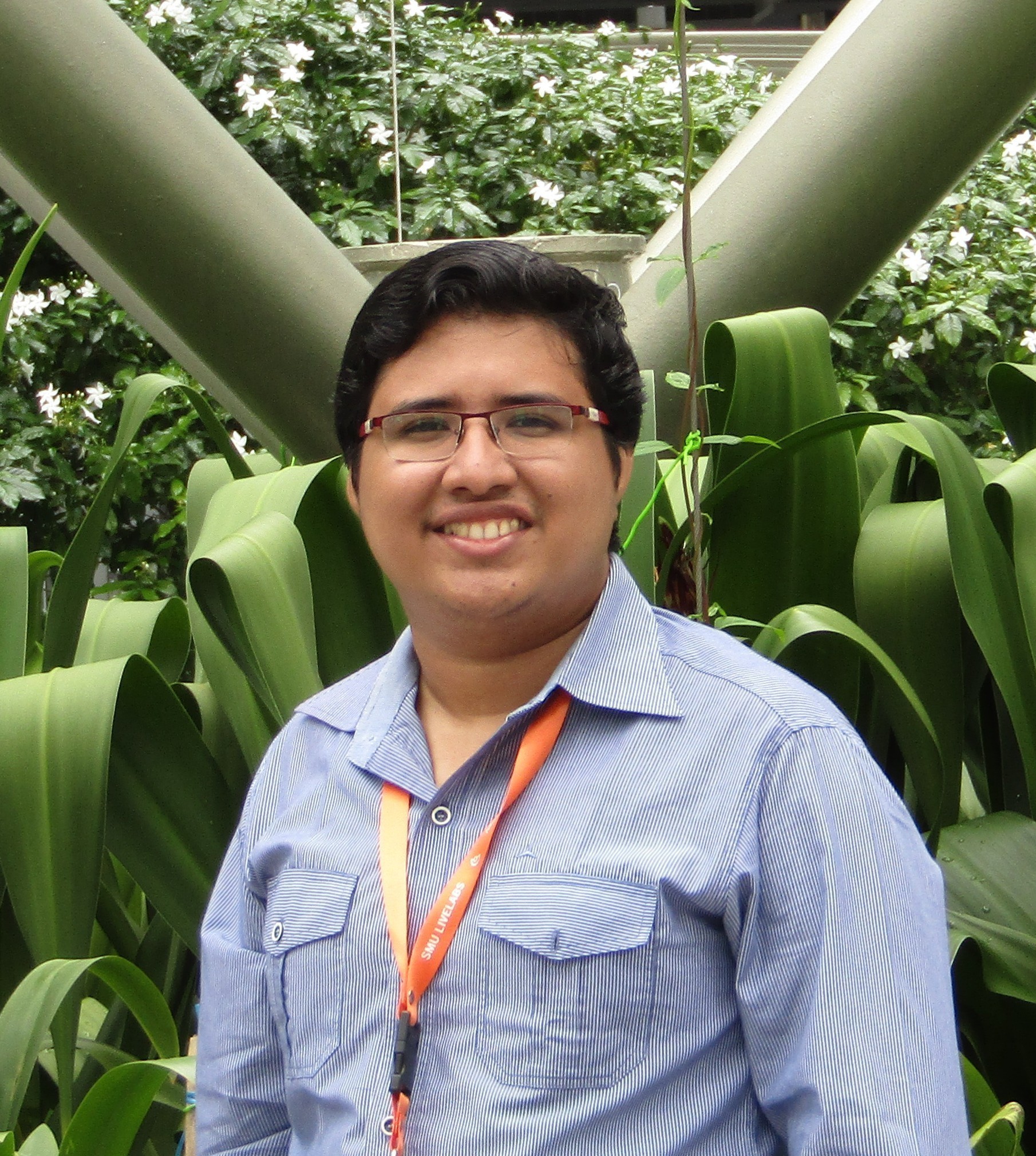}}]%
{Toufique Ahmed}
is a Ph.D. student at UC Davis. He received his B.Sc. and M.Sc. in Computer Science and Engineering from Bangladesh University of Engineering and Technology (BUET) in 2014 and 2016. His research interest includes Software Engineering, the Naturalness of Software, Machine Learning, and Sentiment Analysis. He is the recipient of the five-year prestigious Dean’s Distinguished Graduate Fellowship (DDGF) offered by The Office of graduate studies,  The College of Engineering, and The Graduate Group in Computer Science, UC Davis.

\end{IEEEbiography}

\begin{IEEEbiography}[{\includegraphics[width=1in,height=1.25in,clip,keepaspectratio]{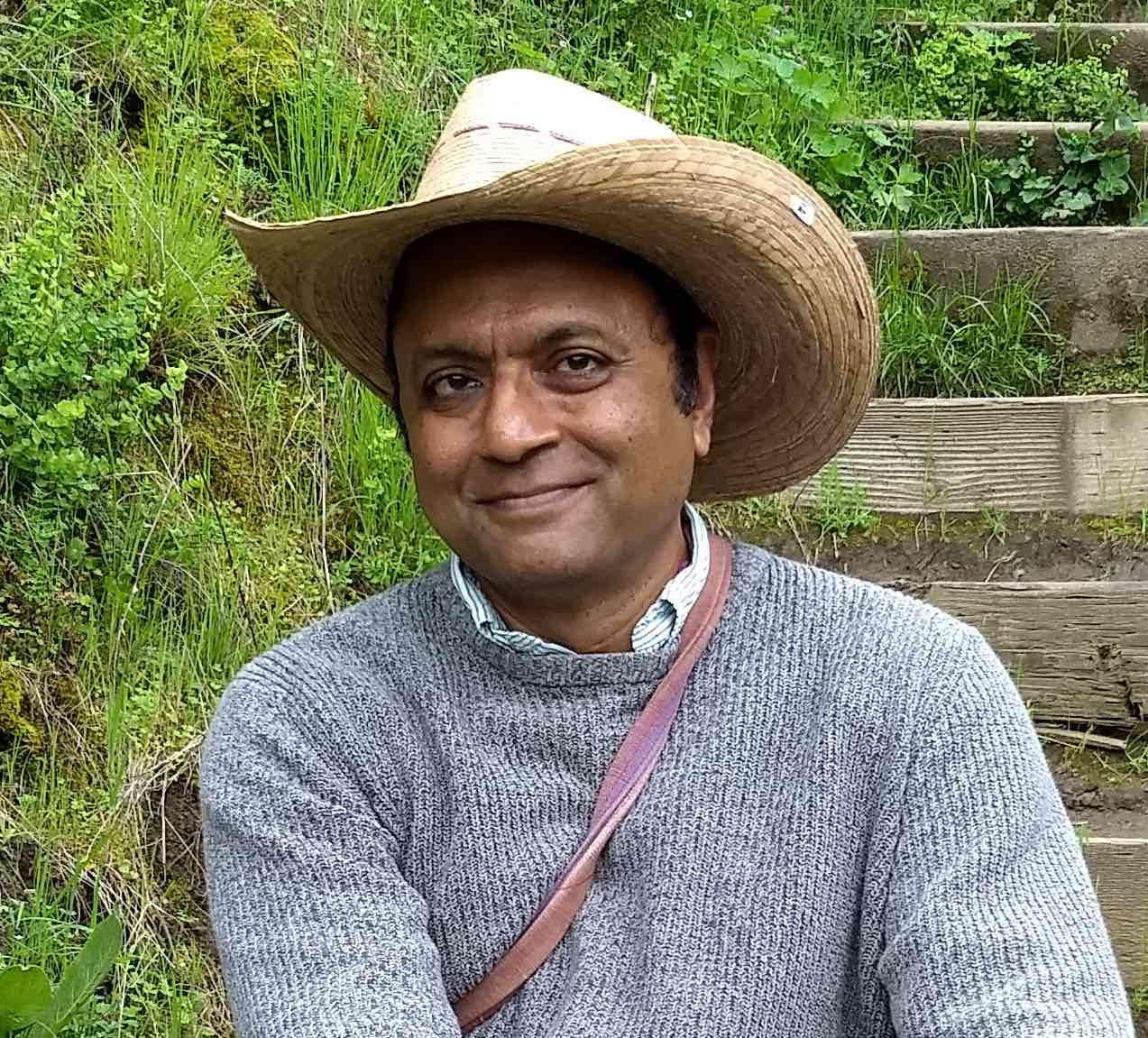}}]%
{Premkumar Devanbu}
received his B. Tech from IIT Madras and his Ph.D from Rutgers University. He is currently Distinguished Professor of Computer Science at UC Davis. His research interests include Empirical Software Engineering and the applications of the Naturalness of Software, including Machine Learning applied to Software Engineering.

\end{IEEEbiography}

\begin{IEEEbiography}[{\includegraphics[width=1in,height=1.25in,clip,keepaspectratio]{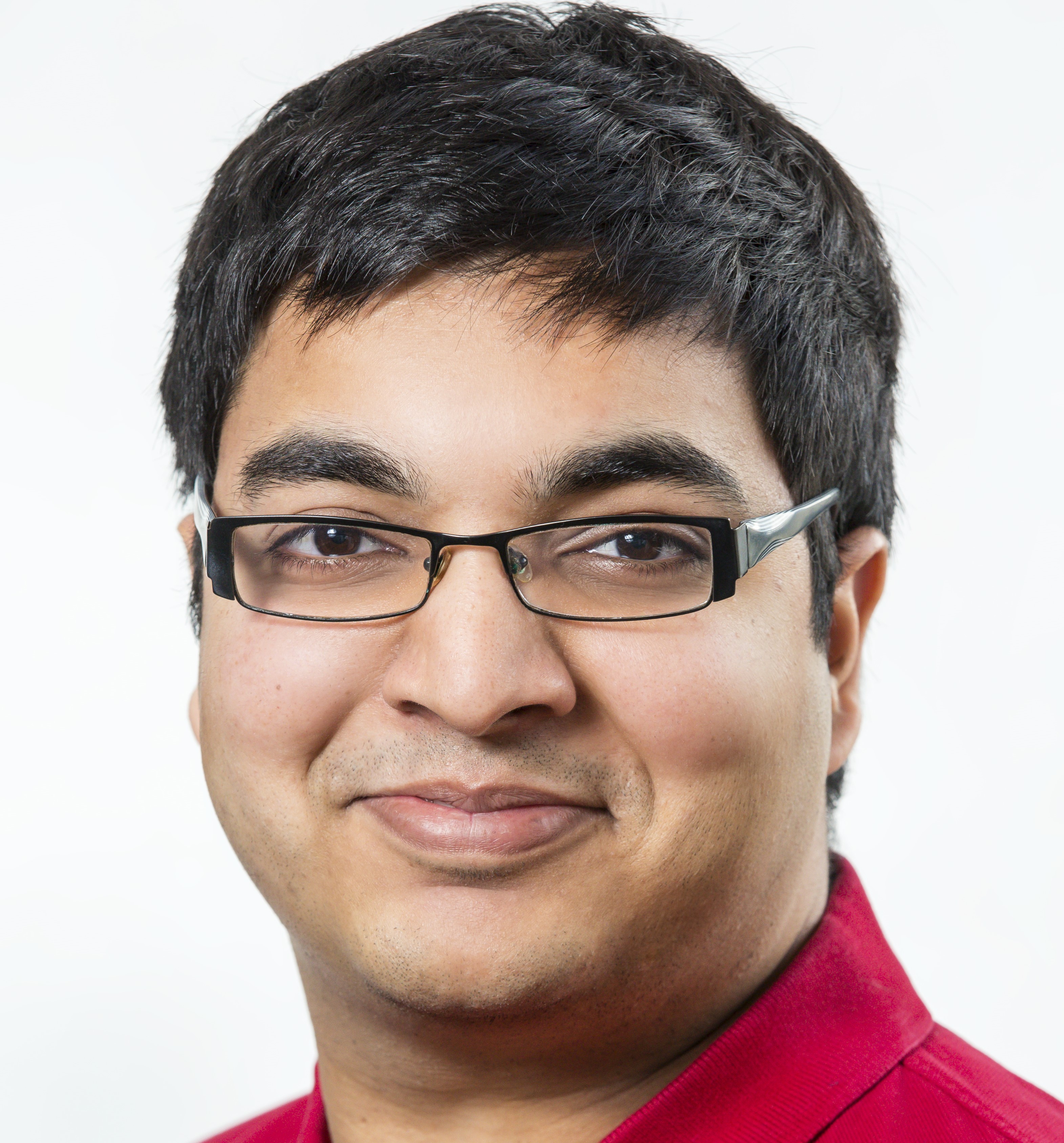}}]
  {Anand Ashok Sawant} is a Postdoctoral scholar working on the application of Machine Learning for Software
  Engineering at the Decal lab at the University of California Davis, USA. He got is PhD at the Delft University of Technology in 2019, where his research was focused on studying API evolution.

\end{IEEEbiography}




\end{document}